\documentclass[sigconf]{acmart}
\usepackage[nameinlink,capitalise]{cleveref}

\newif\iftrack
\trackfalse 

\crefname{table}{Tab.}{Tabs.}
\crefname{table}{Table}{Tables}
\crefname{section}{Sec.}{Secs.}
\crefname{section}{Section}{Sections}
\AtBeginDocument{%
  }

\copyrightyear{2026}
\acmYear{2026}
\setcopyright{cc}
\setcctype{by}
\acmConference[CHI '26]{Proceedings of the 2026 CHI Conference on Human Factors in Computing Systems}{April 13--17, 2026}{Barcelona, Spain}
\acmBooktitle{Proceedings of the 2026 CHI Conference on Human Factors in Computing Systems (CHI '26), April 13--17, 2026, Barcelona, Spain}
\acmPrice{}
\acmDOI{10.1145/3772318.3791122}
\acmISBN{979-8-4007-2278-3/2026/04}

\usepackage{soul}  
\usepackage{color} 

\newcommand{\bstart}[1]{\vspace{1mm} \noindent{\textbf{#1.}}}
\newcommand{\bfstart}[1]{\vspace{1mm} \noindent{\textbf{#1}}}
\newcommand{\istart}[1]{\vspace{1mm} \noindent{\textit{#1.}}}

\newcommand{\matt}[1]{}

\newcommand{\di}[1]{\textbf{D1}}
\newcommand{\dii}[1]{\textbf{D2}}
\newcommand{\diii}[1]{\textbf{D3}}

\newcommand{\usera}[1]{Abhi}
\newcommand{\userd}[1]{David}

\newcommand{\ie}{i.e.,\ }
\newcommand{\etal}{et al.}
\newcommand{\eg}{e.g.,\ }

\begin{document}

\title[Glass Chirolytics]{Glass Chirolytics: Reciprocal Compositing and Shared Gestural Control for Face-to-Face Collaborative Visualization at a Distance}

\author{Dion Barja}
\authornote{Barja performed this work while at the University of Waterloo.}
\orcid{0009-0006-5720-8904}
\affiliation{%
  \institution{University of Manitoba}
  \city{Winnipeg}
  \state{Manitoba}
  \country{Canada}
}
\email{barjad@myumanitoba.ca}

\author{Matthew Brehmer}
\orcid{0000-0001-5524-2291}
\affiliation{%
  \institution{University of Waterloo}
  \city{Waterloo}
  \state{Ontario}
  \country{Canada}}
\email{mbrehmer@uwaterloo.ca}

\renewcommand{\shortauthors}{Barja \& Brehmer}

\begin{teaserfigure}
    \centering
    \includegraphics[width=\textwidth]{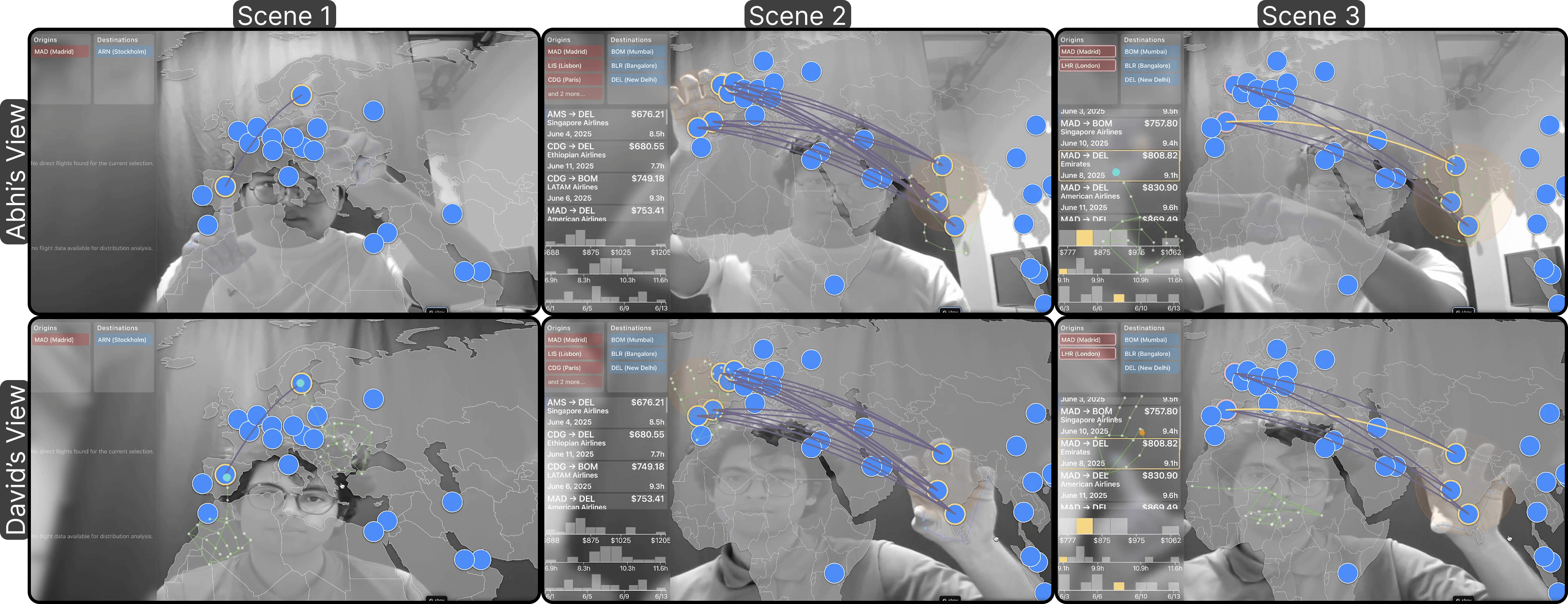}
    \caption{A collaborative visualization interface incorporating our Glass Chirolytics approach, captured from both David and Abhi's perspectives as they jointly decide upon an travel itinerary. In Scene 1, David selects an origin (Madrid) and destination (Stockholm) with their left and right hands, respectively. In Scene 2, David and Abhi both spread a hand to select multiple origin and destination airports in Western Europe and India, respectively. In Scene 3, David scrolls a list of flights with their left hand while pointing at one listed flight with their other hand. Abhi continues to select destinations in India while selecting a listed flight with their left hand.}
    \Description{A gallery of six screenshots within the application using Glass Chirolytics. The six screenshots are organized into two perspectives for each of the three scenes in a two-by-three grid. The two perspectives are of two collaborators, David and Abhi, as they work on a shared, interactive visualization depicting a commercial flight map. In Scene 1, David points with their left and right index finger at two distinct nodes to select an origin and destination, respectively. Abhi watches as David performs this action. In Scene 2, David spreads their left hand and Abhi spreads their right hand to select multiple origins and destinations, respectively. In Scene 3, David makes an 'ok' gesture with their left hand to drag a list of flights and uses their right hand to point at a flight of interest. Abhi spreads their right hand to select a set of destinations and points their left index finger to select the flight David is pointing at.}
  \label{fig:teaser}
\end{teaserfigure}

\begin{abstract}
Videoconference conversations about data often entail screen sharing visualization artifacts, in which nonverbal communication goes largely ignored. Beyond presentation use cases, conversations supported by visualization also arise in collaborative decision making, technical interviews, and tutoring: use cases that benefit from participants being able to see one another as they exchange questions about the data. In this paper, we employ a reciprocal compositing of visualization and interface widgets over the mirrored video of one's conversation partner, suggestive of a pane of glass, in which both parties can simultaneously manipulate composited elements via bimanual gestures. We demonstrate our approach with implementations of several visualization interfaces spanning the aforementioned use cases, and we evaluate our approach in a study (N = 16) comparing it to videoconferencing while using a mouse to interact with a collaborative web application. Our findings suggest that our approach promotes feelings of presence and mutual awareness of analytical intent.

\end{abstract}



\maketitle

\section{Introduction}
\label{sec:intro}

Videoconference applications continue to be a prevalent medium for synchronous remote communication between knowledge workers~\cite{barrero2023evo, holtz2022remote}, communication that often involves the sharing and discussion of data visualization and other analytics artifacts~\cite{brehmer2022jam}.
When using applications such as Microsoft Teams \cite{teams2022} or Zoom \cite{zoom2022}, a single participant assumes a presenter role and screen shares slide presentations or dashboard applications. 
When screen sharing, others cannot interact with these artifacts, and participants' webcam feeds are relegated to thumbnail videos in the periphery. 
As a consequence, participants' attention is split and they lose awareness of nonverbal communication cues omnipresent in face-to-face conversations, such as deictic gestures and eye contact.
Recent work~\cite{hall2022chironomia,femi2024visconductor,liao2022realitytalk} has explored the potential of augmenting a presenter's webcam feed by compositing visual artifacts atop it, which are in turn revealed or highlighted by a presenter's hands gestures. With computer vision-based recognition, these expressive gestures can now also be functional.    
However, by following a strict \textit{what-you-see-is-what-I-see} (WYSIWIS) approach~\cite{stefik1987wysiwis}, this augmentation perpetuates a pattern of one-way communication from presenter to audience and does not speak to the need for nonverbal communication in scenarios \textit{beyond presentation}, such as collaborative ideation and analytical decision-making~\cite{cappelli2021future}.
Moreover, many effective presentations that communicate observations grounded in data~\cite{schwabish2016better} can be supported with conventional statistical charts modified with simple reveal and highlight animations that draw attention to individual observations.
In contrast, open-ended collaborative analytical scenarios may entail a greater breadth and complexity of visualization artifacts, with a commensurately richer set of interactions to modify them~\cite{yi2007toward}.

In this paper, we extend augmented videoconferencing to support rich analytical conversations
with a reciprocated form of WYSIWIS augmentation applied to commodity webcam video.
We call our approach \textit{Glass Chirolytics}, as it suggests a single pane of glass separating the participants: our approach composites visualization artifacts and interface widgets atop the mirrored webcam video of a conversation partner. 
Unlike prior approaches to augmented video, which prioritized one-to-many presentation scenarios, ours prioritizes paired data analysis with inherently relational data and allows for either participant to control composited and visually-interconnected elements with bimanual mid-air gestures.  

Our \textbf{research contributions} are as follows. 
First, we propose \textit{Glass Chirolytics} as a novel approach to supporting \textit{face-to-face} paired analytics conversations between remote participants, combining reciprocal WYSIWIS video compositing with gestural interaction, allowing either participant to manipulate shared visualization artifacts. 
We assemble a bimanual gestural vocabulary for supporting analytical conversations, which features two novel gestures as well as logic for supporting gestures performed simultaneously by two sets of hands. This vocabulary prioritizes the analytically useful actions of navigating coordinate spaces as well as the manipulation and persistent selection of visual elements, so as to reveal connecting relationships between them.
Next, we demonstrate the potential utility of our approach through the implementation of seven application interfaces, spanning decision making, collaborative analysis, one-on-one tutoring, and technical interviewing.
Lastly, we report and reflect on findings from an evaluation with 16 participants, one based around a collaborative decision-making scenario. 

\section{Background \& Related Work}
\label{sec:rw}

Earlier work involving the augmentation of videoconference experiences for delivering presentations about data via gestural interaction had theoretical foundations in the \textit{rhetorical} use of gesture for public speaking and in \textit{communicative} use cases for data visualization. 
In contrast, the foundations of this work draw more from the \textit{conversational} utility of gesture in face-to-face exchanges and in \textit{collaborative} use cases for visualization.

\subsection{The Utility of Hand Gestures in Conversation and in Problem Solving}
\label{sec:rw:gestures}

Voluntary nonverbal communication expressed via facial expression as well as via head and hand movement are essential to face-to-face conversation. 
In particular, hand gestures amplify communicative intentions~\cite{dohen2017coprod, wagner2014gesture}, and in educational contexts, they promote discussion engagement and improve learning outcomes~\cite{turgay2023bodylanguage}. 
Moreover, they benefit listener and speaker alike~\cite{clough2020role}: for the former, gestures provide extra context when speech is unclear, improving overall comprehension, while for the latter, gestures support the production of speech. 
Despite these benefits and a consensus that interpersonal communication is more effective when gestures complement speech~\cite{kang2016handsminds}, our hands are seldom visible in videoconferencing applications: common webcam placements tend to capture only a person's head and upper torso~\cite{bailenson2021nonverbal}. 
As a result, the human mirror neuron system (hMNS), a single overarching system governing both linguistic and gestural communication, is engaged to a lesser extent during videoconference calls than in face-to-face conversations~\cite{dickerson2017role}.
Also less apparent during videoconferencing is cross-brain synchrony~\cite{zhao2023separable}: the neural of alignment activity between people during social interactions and a signifier of coordination and empathy.  

\bfstart{Deictic gestures} in particular serve to direct a listener's attention~\cite{kendon2004gesture}, such as pointing at a location or object.
However, with the convention of screen sharing visual aids in videoconferencing applications, deictic hand gestures, even when they appear in a speaker's video frame, are often ambiguous given the relative placement of speaker video and screen shared content on the viewer's display. 

\bfstart{Non-communicative gesturing}, in the meantime, can assist with problem-solving~\cite{goldin2009gesturing} and learning as these gestures can produce embodied representations of problems and concepts~\cite{jamalian2013gestures}. 
These gestures may therefore be good candidates for manipulating virtual objects in mid-air gestural interfaces, and accordingly we use this justification for some of the gestural vocabulary defined below in~\cref{sec:design:gestures}.

\subsection{Augmenting Video Communication}
\label{sec:rw:augvideoconference}

\matt{begin with Three's Company as preamble here}
Augmenting synchronous video communication between remote peers has long been an area of interest in HCI~\cite{engelbart1968research}, and the lineage of this research includes work dedicated to restoring an awareness of gestural communication (\eg~\cite{tang2010threes}) and other non-verbal cues such as facial expressions (\eg~\cite{ishii1992clearboard}) in collaborative tasks.
In recent years, both commercial tools and research projects have introduced approaches to augmented video experiences with composited visual aids, and beyond making gestures visible, some of this work allows for the manipulation of these visual aids via gestural interaction.

\bstart{Compositing webcam video and visual aids}
Several existing commercial videoconferencing tools can composite screen shared content with a speaker's webcam video, producing either a picture-in-picture experience akin to how visual aids appear over the shoulder of a broadcast news anchor, or an experience in which the speaker's outline is segmented and composited atop the content, akin to a broadcast weather reporter. 
These include virtual camera applications such as OBS~\cite{obs2022} and Airtime~\cite{airtime2025}, operating system features like macOS's Presenter Overlay~\cite{macos2025}, and recent webcam segmentation features in Microsoft Teams \cite{teams2022} and Zoom \cite{zoom2022}. 
These tools assume a designated presenter role, in which a single participant updates their visual aids by interacting with the screen shared application's interface.
A limitation of these approaches is that the use of deictic gestures to direct an audience's attention to elements in the composited visual aids requires coordination and practice, particularly if the content is composited behind the speaker. 
To address this, recent research (\eg~\cite{liao2022realitytalk,hall2022chironomia,urban2023clio}) composites content atop of webcam video, a design choice that we also make in our work.

Looking beyond presentation use cases, recent research has looked to support remote and hybrid videoconference meetings by exploring approaches that composite webcam video from multiple participants with shared content.
Gr\o{}nb\ae{}k~\etal's MirrorBlender~\cite{gronbaek2021mirror} and Mirrorverse \cite{gronbaek2023mirrorverse}, for instance, incorporate a \textit{what-you-see-is-what-I-see} (WYSIWIS)~\cite{stefik1987wysiwis} composited experience wherein all participants have a common perspective on the shared content; each participant can manipulate the position, scale, and translucency of their webcam video feeds, thereby adapting the interface in real-time to diverse meeting situations and allowing them to direct the attention of their peers.
Hu~\etal's OpenMic~\cite{hu2023openmic} is another point in this design space, one that applies proxemic metaphors to conversational turn-taking in multi-party teleconference meetings.
In our work, we employ a reciprocal WYSIWIS experience to support face-to-face meetings between two people.

\bstart{Gestural interaction with composited visual aids}
Several of the aforementioned commercial tools (\eg~\cite{macos2025,zoom2022}) also incorporate computer vision and specifically basic pose recognition to trigger animations when a participant performs a static gesture, such as by raising a hand or giving a thumbs-up.
For continuous hand tracking and the recognition of dynamic mid-air hand gestures, depth cameras such as the Microsoft Kinect~\cite{shotton2013real} can be used to reveal and animate visuals composited in the video foreground~\cite{saquib2019interactive}.
More recently, researchers have used frameworks such as OpenPose~\cite{cao2019openpose} and MediaPipe~\cite{valentin2020mediapipe} to achieve similar results with commodity webcams~\cite{cao2024elastica,liao2022realitytalk}. 
These developments have led to augmented video presentation experiences that allow presenters to directly manipulate individual multimedia objects~\cite{urban2023clio} and even the contents of entire web browser tabs composited behind or in front of a presenter~\cite{cordeil2025presenter}, thereby replacing indirect manipulation with pointing devices.  
Thus far, interfaces that composite webcam video with visual aids grant the gestural manipulation of these aids to a single presenter, meaning that when these interfaces are used in videoconferencing scenarios, only one participant can reveal, move, or draw content. 
Moreover, the gestural vocabulary of these interfaces, while likely appropriate for manipulating multimedia and web content, is unlikely to support the breadth and complexity of interaction typical of visual data analysis applications~\cite{yi2007toward}.

\subsection{Collaborative Visualization}
\label{sec:rw:collabvis}

Isenberg~\etal~\cite{isenberg2011collaborative} define \textit{collaborative visualization} as ``\textit{the shared use of computer-supported, (interactive,) visual representations of data by more than one person with the common goal of contribution to joint information processing activities.}'' 
The parenthetical \textit{`interactive'} is notable in this definition, as it serves to distinguish typical presentation scenarios from analytical ones; in the former category, it is likely that a single presenter interacts with visual representations to satisfy the information processing needs of an audience, whereas in the latter category, there is arguably a more equitable access to interactive affordances.
The collaborative visualization design space can be characterized by the two established axes of computer-supported cooperative work~\cite{johansen1988groupware}: \textit{space} (co-located vs. distributed) and \textit{time} (asynchronous vs. synchronous). 
Brehmer and Kosara~\cite{brehmer2022jam} further characterize synchronous collaboration with and around visualization artifacts along a spectrum of \textit{formality}, from informal data analysis sessions between peers to semi-improvised briefings and formal presentations with designated presenter roles. 
We therefore position our work within the \textit{distributed} (\ie~remote or hybrid) and \textit{synchronous} quadrant of this design space, focusing on \textit{informal} data analysis scenarios involving a pair of collaborators.

\bstart{Infrastructure for collaborative visualization}
Early work in remote and synchronous collaborative visualization~\cite{balakrishnan2008syncvis} showed  that the ability for remote peers to concurrently see and manipulate a shared interactive visualization artifact can improve group performance on a collaborative problem-solving task. 
More recently, infrastructural support for remote and synchronous collaborative visualization
includes Badam~\etal's Vistrates~\cite{badam2018vistrates}, which presents a workflow for creating, interacting with, and presenting visualization artifacts for distributed peers, and Schwab~\etal's VisConnect~\cite{schwab2020visconnect}, which supports synchronizing low-level interaction events when many peers interact with these artifacts.
Infrastructure for synchronous collaboration is also appearing in commercial tools, such as Observable's Canvases~\cite{observablecanvases}.
Complementing these projects are Neogy~\etal's~design space~\cite{neogy2020representing} for representing remote peers' concurrent interactions, as well as Han and Isaacs's deictic approach~\cite{han2024deixis} to cursor-based gestural interaction for annotating and drawing peers' attention to visualization elements.
In general, these frameworks and techniques are agnostic as to whether remote collaborators can also be seen and heard, such as via a separate videoconference application. 

\bstart{Collaborative visualization beyond the desktop / beyond mouse and keyboard}
While much of the extant work in remote and synchronous collaborative visualization to date addresses keyboard and mouse interaction with web-based visualization, some explicitly addresses asymmetric interaction and display modalities, such as Tong~\etal~\cite{tong2023asymmetric}'s recent finding that collaborators with asymmetric device capabilities (\ie desktop and head-mounted displays) can achieve similar task performance relative to those with symmetric device capabilities.
However, irrespective of visualization display and interaction modalities and the possible asymmetries between collaborators, prior synchronous and remote collaborative visualization does not directly acknowledge the value of being able to see (or hear) one's collaborators and the nonverbal communication cues they make while performing a collaborative task, a value that is central to our current work.
A recent exception is Borowski~\etal's DashSpace~\cite{borowski2025dashspace}, which enables synchronous collaborative immersive analytics by those using either head-mounted displays or desktop devices; those using the former appear as silhouette avatars and those using the latter appear as floating video thumbnails in the periphery of a 3D scene, both serving to provide a sense of collaborator presence.
In our work, we concentrate on low-cost commodity webcams rather than head-mounted displays, and by compositing visualization artifacts with collaborator video, our goal is not only presence, but also to promote the reception of nonverbal communication that a face-to-face perspective offers.

\subsection{Visualization in Augmented Video Presentations}
\label{sec:rw:augvis}

In recent years, several research projects~\cite{hall2022chironomia,femi2024visconductor,kristanto2023hanstreamer,takahira2025insitutale} have demonstrated gestural interaction with composited visualization artifacts for video-based presentations, each incorporating a WYSIWIS approach in which both a presenter and their audience see artifacts composited over the presenter's webcam video.
The common inspiration for these tools can be traced to a 2010 BBC documentary hosted by the late Hans Rosling~\cite{bbc2021youtube}, in which a semi-transparent animated bubble chart appears in the foreground; meanwhile, Rosling's gesticulations give the impression that he is controlling the dynamic chart animation with his movements. 
The types of visualization artifacts and gestural vocabulary featured in these projects understandably prioritize presentation scenarios. 
First, Hall~\etal~\cite{hall2022chironomia} demonstrated bimanual rhetorical gestures for revealing, annotating, and comparing elements in common statistical charts such as bar, line, and proportion charts. 
Femi-Gege~\etal's VisConductor~\cite{femi2024visconductor} then extended this vocabulary with continuous gestures for controlling playback in animated bubble and rank bar charts, however this more complex vocabulary necessitated the use of a secondary presenter display annotated with gestural hints and detection feedback. 
Most recently, Takahira~\etal's InSituTale~\cite{takahira2025insitutale} is also reminiscent of Rosling with the integration of physical objects in presentations about data, with gestures that manipulate common household objects that trigger updates to composited visualization artifacts based on the position and orientation of the held objects. 
In this work, we explicitly depart from presentation use cases~\cite{hall2022chironomia,femi2024visconductor,takahira2025insitutale} with an augmented video approach for addressing collaborative data analysis scenarios. 
Our contribution extends the design space of this emerging medium by incorporating a gestural vocabulary grounded in an interaction taxonomy for visual analysis~\cite{yi2007toward}, with composited visualization artifacts reflecting complex spatial and relational structures, such as node-link network graphs and origin-destination trajectory maps.
\section{The Design of Glass Chirolytics}
\label{sec:design}

\begin{figure*}[h]
    \centering
    \includegraphics[width=\textwidth]{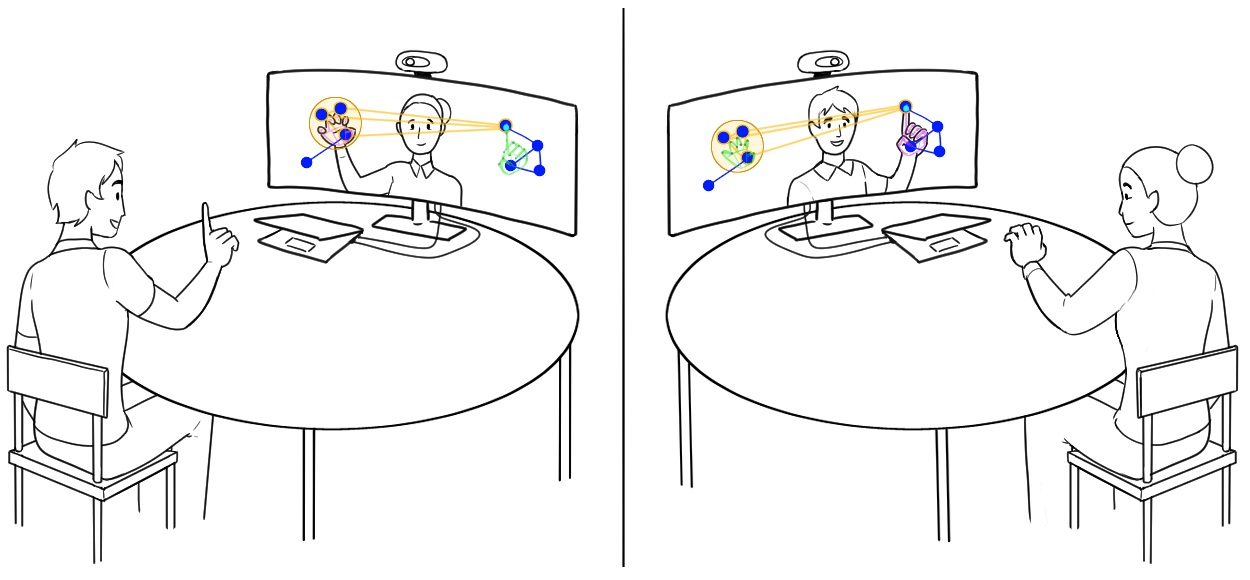}
    \caption{In this illustrated instance of the Glass Chirolytics approach, a remote collaborator appears composited behind a synchronized node-link diagram. The collaborator on the left points with their right hand to select one node while the collaborator on the right selects multiple nodes with their left hand; both can see a reflection of their hands as a green skeletal mesh composited in their local foreground. Lastly, the combination of their gestures triggers the reveal of highlighted links between the set of selected nodes.}
    \Description{A side-by-side illustration of two users interacting with the Glass Chirolytics system from a physical perspective. Each user is seated at a desk with a laptop connected to a monitor that displays the application. A standard webcam is mounted on top of the monitor, capturing their hand gestures for interaction. The user on the left is shown using their right index finger to point and select a single node within a cluster on the right side of the visualization. The user on the right, meanwhile, uses a spread-hand gesture with their left hand to select three nodes within a cluster on the left side of the visualization.}
    \label{fig:diagram}
\end{figure*}

We propose \textit{Glass Chirolytics} (\cref{fig:diagram}) as an approach to face-to-face pair analytics~\cite{arias2011pair} between two remote collaborators, an approach developed for the medium of gesture-aware augmented videoconferencing.
The name combines the metaphor of a glass partition between participants with a neologism referencing the use of our hands (\textit{`chiro–')} for collaboration around shared visual \textit{analytics} artifacts. 

\subsection{Key Design Decisions}
\label{sec:design:goals}

\begin{figure*}[t]
    \centering
    \includegraphics[width=\textwidth]{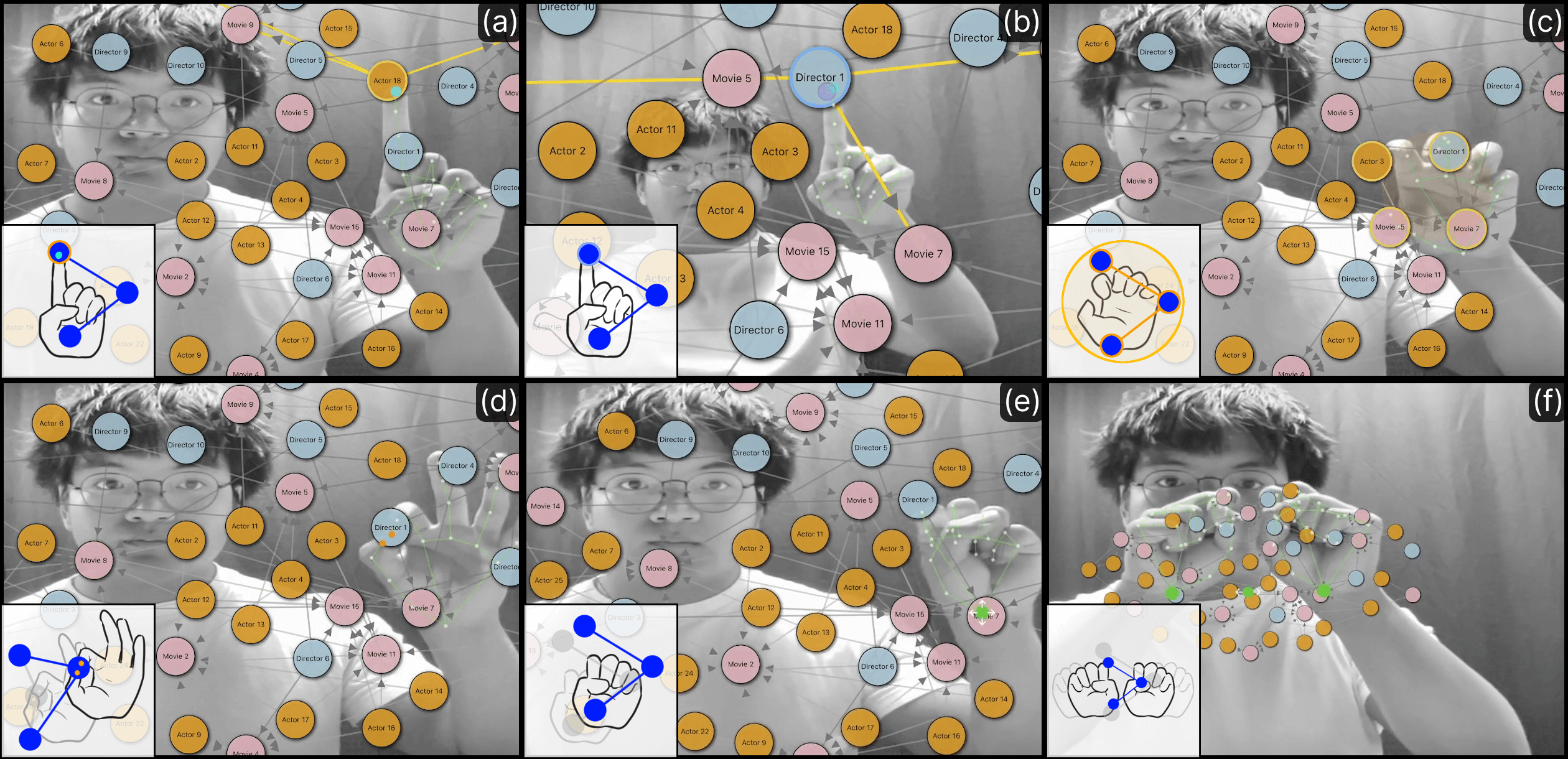}
    \caption{Our gestural vocabulary includes: (a) \textit{pointing} to ephemerally select an individual element; (b) \textit{point-and-tap} to persist an individual selection; (c) \textit{spread} to coarsely select multiple elements (ephemeral unless held for one second); (d) \textit{pinch-and-move} to reconfigure the positions of elements; (e) \textit{grab-and-move} panning; and (f) \textit{separate-or-join} zooming. Recognition of the latter two gestures follows a one-second delay communicated with an ephemeral green indicator icon appearing at the base of the wrists.}
    \Description{A gallery of six screenshots within the application using Glass Chirolytics labeled (a) to (f) displaying six different interaction techniques performed by a user. Within each screenshot is an inset illustration of the interaction technique. Image (a) shows the user pointing with their right index finger to select a node. Image (b) shows them making a similar pointing gesture, but after extending their thumb from the initial position. Image (c) shows them spreading their right hand, forming a circle around the tip of their fingers. Nodes within the circle are selected. Image (d) shows them maintaining an 'ok' gesture to move a node. Image (e) shows them maintaining a fist with their right hand to pan the visualization. Image (f) shows them holding a fist with both hands to enable them to zoom in or out of the visualization.}
    \label{fig:vocab}
\end{figure*}

Informed by prior work in augmented video (Sections~\ref{sec:rw:augvideoconference}, \ref{sec:rw:augvis}), we made three governing design decisions:

\bstart{\di{}: Provide a common perspective on shared visual aids while coordinating face-to-face conversation}
We depart from a strict \textit{what-you-see-is-what-I-see} (WYSIWIS) approach~\cite{stefik1987wysiwis} in which all participants see the same interface. 
Instead, by focusing on the give and take of analytical conversations between two people, we ensure that one can always see their conversation partner, rather than a reflection of themselves.
While this mutual visibility is also achieved in conventional videoconferencing, as soon as those conversations require visual aids, this visibility is compromised with the shift to screen sharing.
We therefore composite visual aids over the webcam video of one's conversation partner, with the latter shown in grayscale to boost the salience of the visual aids (Figures~\ref{fig:teaser}, \ref{fig:vocab} --- \ref{fig:apps:interview}).
The effect is not unlike two people facing one another with a pane of glass or a semi-transparent display material~\cite{gong2023side} separating them. 
However, an issue that arises when using these display materials in a co-located setting is that content must be legible from both sides; in other words, it must be symmetric, which precludes the display of text and many visualization design conventions. 
In remote settings, this issue is resolved simply by compositing content over a horizontally-mirrored video feed of one's conversation partner, which ensures that both participants are looking (and gesturing) at the same visual aids. 

While we do away with the `self-view' video of conventional videoconferencing, we nevertheless provide both participants with visual feedback (\cref{fig:diagram}). Composited atop the visual aids is feedback for the local collaborator's hands, appearing as a bright green skeletal mesh, while the successful recognition of their gestures appears as ephemeral activation icons.
Altogether, we characterize our approach to augmented videoconferencing as employing a \textit{reciprocal compositing} of remote video, shared visual aids, and gestural feedback.

\bstart{\dii{}: Support the collaborative analysis of complex data abstractions}
Prior work exploring the use of augmented video to talk about data with remote audiences (\cref{sec:rw:augvis}) understandably prioritized presentation-oriented visualization idioms~\cite{kosara2016presentation}. 
This includes a palette of familiar statistical charts for presenting observations about tabular data~\cite{hall2022chironomia} or animated approaches for suspenseful storytelling about time-varying data~\cite{femi2024visconductor}. 
However, analytical use cases involving visualization~\cite{munzner2014visualization} often entail a targeted interest beyond individual values, attribute distributions, or trends, and towards more complex structural constructs such as correlations, (spatial) relationships, and topologies.
We therefore prioritize the categories of visualization artifacts associated with these higher-level abstractions, such as origin-destination network maps (\cref{fig:teaser} and ~\cref{fig:apps:eda}: Scene 2) and node-link diagrams (\cref{fig:vocab} and ~\cref{fig:apps:tutoring}: Scene 1).  

\bstart{\diii{}: Synchronize interaction with a gestural vocabulary commensurate with analytical use cases} 
In data presentation scenarios, seldom do presenters need to depart from showing and comparing values in common statistical charts, manipulated through progressive reveal and ephemeral highlighting~\cite{brehmer2022jam}.
Cordeil~\etal's Hanstreamer~\cite{kristanto2023hanstreamer} is therefore notable for its showcasing of dense node-link diagrams along with a gestural vocabulary incorporating persistent selection, zooming, and the repositioning of visual elements, suggesting that this medium could accommodate more complex analytical use cases.
Our gestural vocabulary, characterized below in \cref{sec:design:gestures}, allows two people to concurrently ask questions of the data, questions that require a broader palette of interactions than value highlighting and pairwise value comparisons~\cite{yi2007toward}.
Given our aforementioned interest in structural data abstractions (\dii{}), our gestures can correspond with structural elements at varying levels of specificity: a single element, a set of elements, or multiple sets.
As our approach synchronizes interaction and the state of the composited visual aids, each participant can contribute one or both of their hands to the interaction.

\subsection{Gestural Vocabulary}
\label{sec:design:gestures}

To realize \diii{}, we began by considering common mouse and keyboard actions typical in interactive exploratory data analysis tools and reflected in taxonomies of interaction for information visualization~\cite{yi2007toward}: clicking or lassoing to \textit{select} elements, scrolling or clicking and dragging on navigation widgets to \textit{explore} the spatial distribution of elements, clicking elements and dragging them to \textit{reconfigure} their placements, clicking on interface widgets to conditionally \textit{filter} elements, and clicking while pressing a modifier key to \textit{connect} elements. Unlike prior work in augmented video presentation~\cite{hall2022chironomia, femi2024visconductor} prioritizing rhetorical flourish and allowing for ambiguity, we sought a vocabulary that would prioritize functional clarity to differentiate these five actions. 
We also required gestures that could be easily distinguished from commodity webcam video input.

\bstart{\textit{Point}: ephemeral fine \textit{selection} of an individual element (\cref{fig:vocab}a)}
Pointing the index finger at a single visual element will ephemerally highlight it, such as by modifying the stroke, drop shadow, or fill color of an element in a node-link network representation, or by showing a text label. As this deictic gesture primarily serves to quickly draw the attention of one's conversation partner, moving the finger away will revert the element's appearance.
We adopt this gesture from prior work in augmented videoconferencing~\cite{hall2022chironomia, kristanto2023hanstreamer}.

\bstart{\textit{Point-and-tap}: persistent fine \textit{selection} of an individual element (\cref{fig:vocab}c)}
In the first of two novel gestures introduced in this work, we detect a quick transition away from and immediately back to the fine selection gesture (\cref{fig:vocab}a) by briefly extending the thumb. This is analogous to a click event performed with the thumb when using a mouse with a side button. We indicate that a persistent selection has occurred with a different treatment of an element's visual attributes, such as by applying a different stroke color relative to ephemeral selection. This is also the gesture we employ to interact with composited action widgets emulating buttons or switchers, which we illustrate in \cref{fig:apps:eda} with the selection of a \textit{filter} switcher.

In our initial design of a persistent selection gesture, we sought to replicate a click of the primary mouse button under the index finger or a single finger tap on a touchscreen interface.
However, when facing a standard webcam, this gesture would map to a small motion of the finger briefly toward the camera and immediately recoiling from it. 
In other words, tapping toward the camera exhibited the Heisenberg effect \cite{bowman2001heisenberg}, in which the finger motion changed the position of the selection. 

\bstart{\textit{Spread}: ephemeral and persistent coarse \textit{selection} of multiple elements (\cref{fig:vocab}b)}
Our second novel gesture is detected by spreading all the fingers with one's palm facing the camera, which will trigger a circular selection with a diameter mapped to the extent of the spread, a diameter that will continuously change as the fingers spread or contract and as the distance of the hand from the camera changes. 
This gesture is analogous to a wedge-shaped gesture introduced by Xia~\etal~\cite{xia2017writlarge} for coarse selections on large touchscreen displays, in which the dynamic aperture between the thumb and forefinger determines the size of the projected selection. In our case, this dynamic projection extends toward the camera rather than parallel to a 2D display. When we detect this hand shape, we visually indicate the selection area with a subtle orange circular marquee, and if visualization elements intersect with this marquee, they are ephemerally highlighted as they would be with fine selection until the hand moves away or the hand shape changes.
If, however, one maintains this hand shape for longer than a second, we infer that this held pose reflects a deliberate intent to persist the selection of any elements within the circular marquee.

\bstart{Combinations of coarse and fine selection gestures for two sets of hands} While the three gestures described thus far are single-handed, they can be combined in various ways. 
Between two collaborators, there are four hands in play, meaning that there can be up to four fine or coarse selections, and each selection can be ephemeral or persistent. 
Multiple selections can, for instance, \textit{connect} selected elements by highlighting links between them (\cref{fig:teaser}, \cref{fig:diagram}). 
Moreover, since it is possible to determine which hand (left or right, local or remote) is making a selection, applications employing our approach can use this information in their selection logic, which we demonstrate in two scenarios (Sections~\ref{sec:apps:decision} and~\ref{sec:apps:analysis}).

\bstart{\textit{Pinch-and-move}: \textit{reconfigure} the positions of individual elements (\cref{fig:vocab}d)}
Pinching the index finger and thumb together allows one to `grab' a visual element and reposition it by moving the hand while maintaining the pinch. As with point-and-tap, the detection of this gesture needs to be visible when facing the camera, so the orientation of the pinch must be parallel to the display, recalling the `OK' gesture. 
Like our ephemeral pointing gesture, we adopt this gesture from prior work in augmented videoconferencing~\cite{hall2022chironomia}.
Meanwhile, we associate a two-handed \textit{pinch-and-hold} variant of this gesture with \textit{connecting} elements, such as by drawing a connection between elements if one did not exist previously. 

\bstart{\textit{Grab-and-move}: panning to \textit{explore} the spatial distribution of elements (\cref{fig:vocab}e)}
Forming and holding a fist allows one to `grab' the underlying coordinate space on which elements are drawn and shift it vertically or horizontally. Visual feedback in the form of a circular green progress icon at the base of the wrist (visible only to who is performing the gesture) indicates a recognition of a deliberate intent to pan; releasing the fist disables `pan' mode. We adopt this gesture from Hosseini~\etal's consensus survey of mid-air gestures across application domains~\cite{hosseini2023gestureset}.

\bstart{\textit{Separate-or-join}: zooming to \textit{explore} the spatial distribution of elements (\cref{fig:vocab}f)}
Finally, forming and holding a fist with both hands triggers a `zoom' mode, in which separating the fists zooms in, while bringing the fists together zooms out, analogous to a touchscreen zooming. The visual feedback to enter `zoom' mode is similar to when panning, and likewise releasing either fist disables the mode. Like panning, we adopt this gesture from Hosseini~\etal's survey~\cite{hosseini2023gestureset}.

While \textit{point}, \textit{spread}, and \textit{pinch-and-move} are simultaneously deictic and functional, \textit{grab-and-move} and \textit{separate-or-join} are non-communicative according to the classification discussed above in \cref{sec:rw:gestures}.
However, the latter two may yet be useful for a remote collaborator to witness, particularly if both collaborators have an intent to explore the spatial distribution of elements.

\subsection{Implementation Details}
\label{sec:design:implementation}

Our approach uses Yjs~\cite{yjs}, a conflict-free replicated datatype (CRDT), to sync the underlying data for the composited visualization elements, synchronizing a shared state document used to render both collaborators' views. 
WebRTC~\cite{webrtc} synchronizes this shared state document, as it needs to be updated in real time, in parallel to the remote video feeds. 
We use commodity webcam video to track hand shape and movement.
Our gesture classification pipeline starts with the MediaPipe~\cite{valentin2020mediapipe} hand landmark detection model, identifying the hand and its key points. 
The outputs of that model are the inputs for the MediaPipe hand gesture classification model, which identifies the gesture of a given hand. 
We trained a custom hand gesture classification model by combining of a subset of the HaGRID hand gesture dataset~\cite{kapitanov2024hagrid} and our own training data.
Finally, we built the interfaces featured throughout this paper and in our supplemental video  in React~\cite{react}, and we used D3.js to generate the SVG-based visualization elements~\cite{bostock2011d3}. 
Our implementation is available under an open-source license at \href{https://github.com/ubixgroup/Glass-Chirolytics}{github.com/ubixgroup/Glass-Chirolytics}.

\section{Application Scenarios}
\label{sec:apps}

\begin{figure*}[h]
    \centering
    \includegraphics[width=\textwidth]{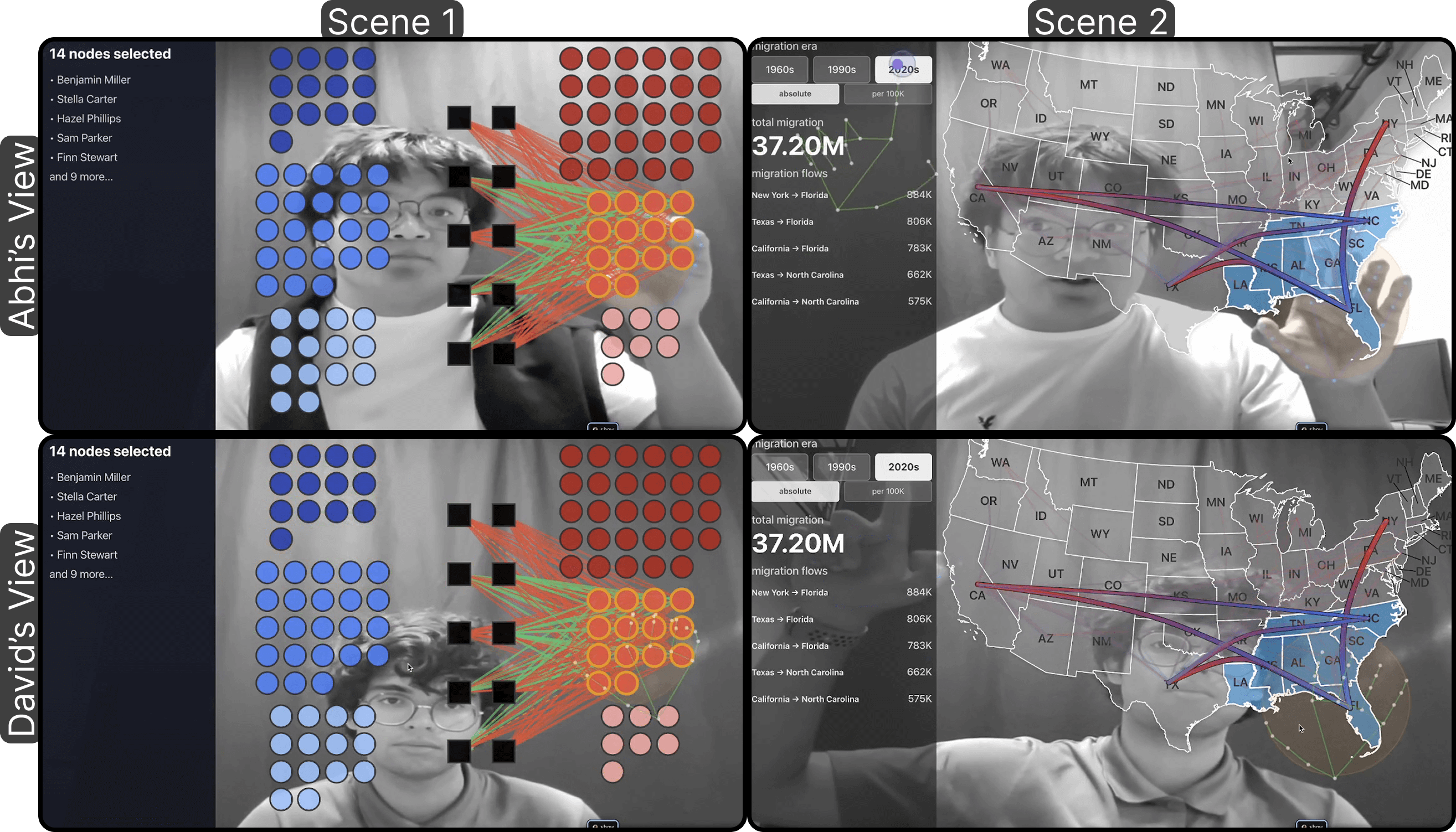}
    \caption{\textit{Exploratory analysis} scenarios: in Scene 1, David spreads their hand to select a group of establishment legislators on the right, revealing their voting patterns. In Scene 2, David spreads their right hand to select states in the American Southeast, revealing the largest sources of incoming domestic migration to those states, while Abhi filters the data with their left hand to reflect the 2020s.}
    \Description{A gallery of four screenshots within the application using Glass Chirolytics. The four screenshots are organized into two perspectives for two scenes in a two-by-two grid. The two perspectives are of two collaborators, David and Abhi, as they work on a shared, interactive visualization. Scene 1 depicts a visualization consisting of elected officials and bills they have voted on. David spreads their right hand to select a set of nodes that represent a political bloc of senators as Abhi watches. Scene 2 depicts a interactive visualization of domestic migration in the United States in which states can be selected to see in-migration and out-migration for the selected states. David spreads their right hand to select several states in the American Southeast as a set of destinations for a query. Abhi performs a point-and-tap to display domestic migration data for the 2020's.}
    \label{fig:apps:eda}
\end{figure*}

\begin{figure*}[h]
    \centering
    \includegraphics[width=\textwidth]{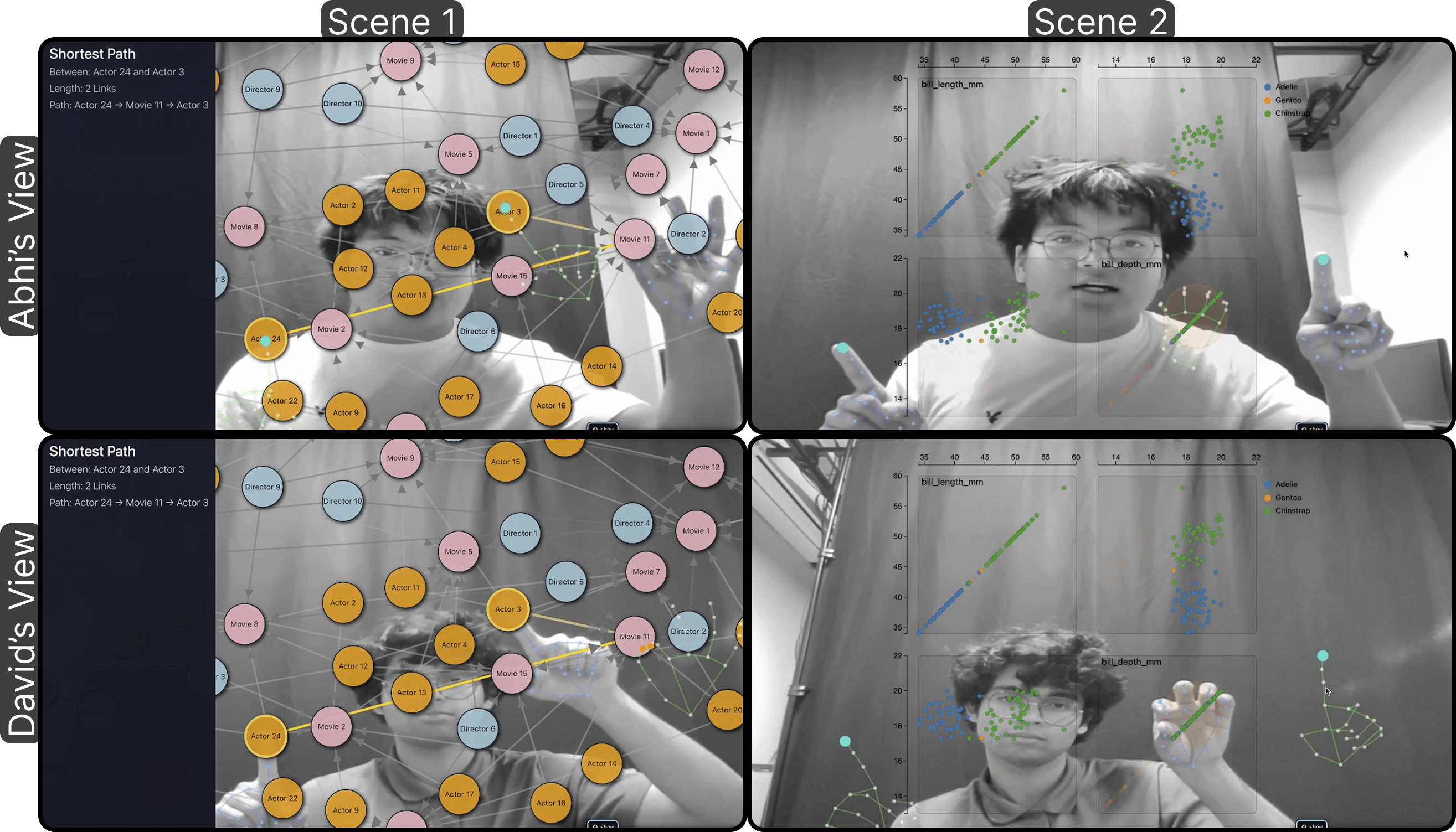}
    \caption{\textit{Tutoring} scenarios: In Scene 1, Abhi points to different nodes with both hands to reveal the shortest path between them, while David moves the path's intermediary node a new position via \textit{pinch-and-move}. In Scene 2, Abhi spreads their right hand to select points in the bottom-right plot of a SPLOM, revealing the selected data points' positions in the other plots.}
    \Description{A gallery of four screenshots within the application using Glass Chirolytics. The four screenshots are organized into two perspectives for two scenes in a two-by-two grid. The two perspectives are of two collaborators, David and Abhi, as they work on a shared, interactive visualization. Scene 1 depicts a node-link diagram of actors, movies, and directors. David maintains an 'ok' gesture with their right hand on a node to drag it while Abhi points with both index finger at distinct nodes. Scene 2 depicts a two-by-two scatterplot matrix depicting data on penguins. David watches as Abhi spreads their right hand to select a set of points on the bottom right scatterplot.}
    \label{fig:apps:tutoring}
\end{figure*}

\begin{figure*}[h]
    \centering
    \includegraphics[width=\textwidth]{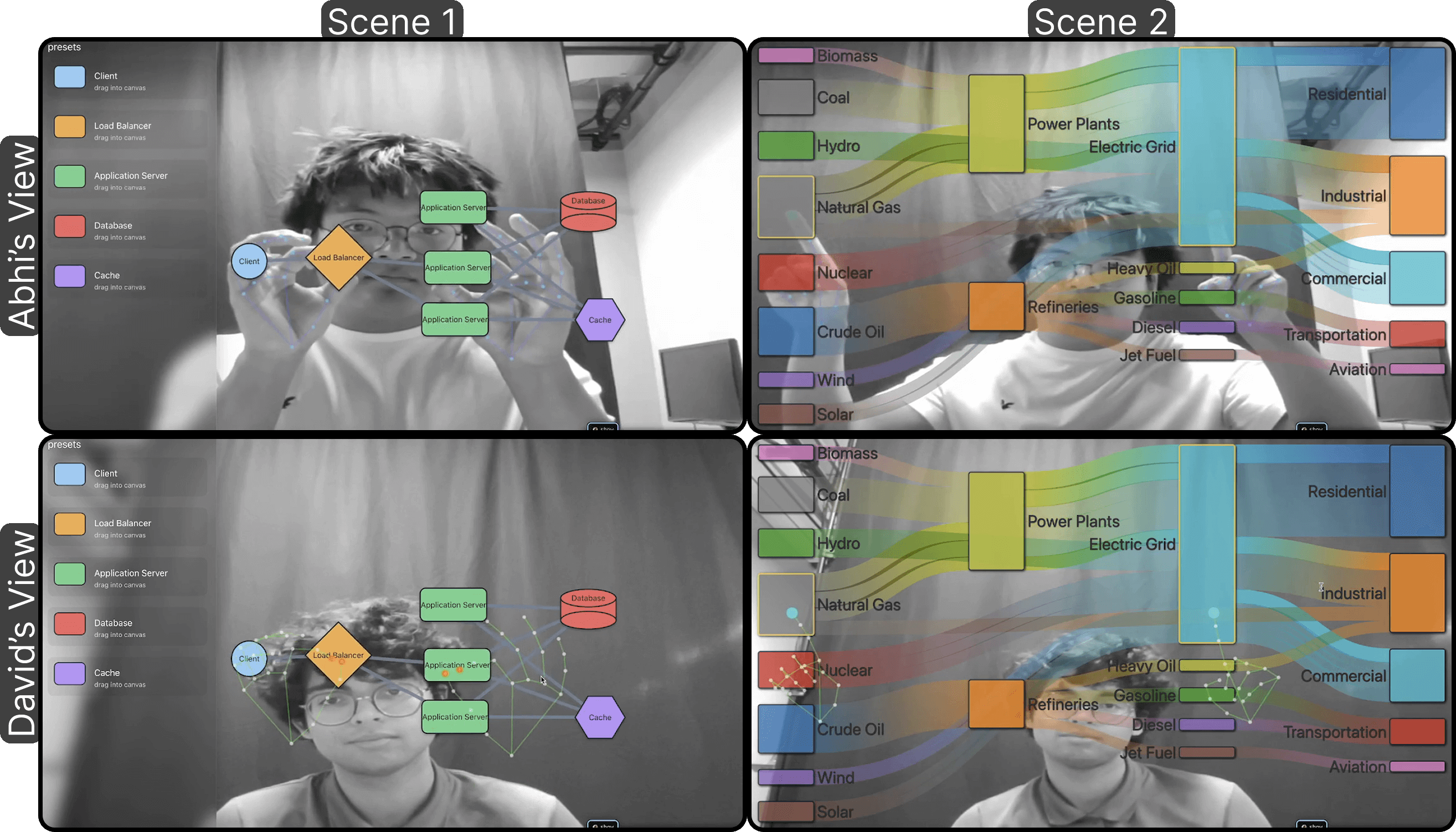}
    \caption{\textit{Technical interviewing} scenarios: In Scene 1, David repositions two elements in a system design diagram. In Scene 2, David points at two nodes in a Sankey energy diagram to highlight the flow of natural gas to the electric grid.}
    \Description{A gallery of four screenshots within the application using Glass Chirolytics. The four screenshots are organized into two perspectives for two scenes in a two-by-two grid. The two perspectives are of two collaborators, David and Abhi, as they work on a shared, interactive visualization. Scene 1 depicts a whiteboard interface in which preset system design assets can be placed inside of. David maintains an 'ok' gesture with the left and right hand on a load balancer and application server asset, respectively. Scene 2 depicts a Sankey energy diagram which displays how energy sources flow into intermediaries and end-users. David points to 'Natural Gas' and 'Electric Grid' with their left and right hands, respectively.}
    \label{fig:apps:interview}
\end{figure*}

We implemented seven visualization interfaces that demonstrate the depth and breadth of the Glass Chirolytics approach across four application scenarios: \textit{decision making}, \textit{exploratory analysis}, \textit{tutoring}, and \textit{technical interviewing}, exhibiting a variety of structural data abstractions and corresponding analytical visualization idioms (\dii{}). 
These interfaces appear in Figures~\ref{fig:teaser},~\ref{fig:apps:eda} --- ~\ref{fig:apps:interview}, and in the supplemental video. 
Throughout these demonstrations, actors \usera{} and \userd{} use the interfaces and execute the gestures in our vocabulary (\diii{}). 
We discuss the \textit{decision making} scenario (\cref{fig:teaser}) in greater depth, as this scenario also forms the basis for our study described below in \cref{sec:evaluation}.
For each scenario, we assume that any requisite data is prepared and visualization design and implementation is complete. In other words, the scope of our approach is the set of synchronous collaboration and communication episodes taking place within a longer data analysis and decision-making life cycle. This life cycle also includes work performed asynchronously and individually, from data cleaning and visualization implementation to confirmatory data analysis and documentation.

\subsection{Decision Making}
\label{sec:apps:decision}
People use visualization artifacts as decision aids~\cite{dimara2022decisions} when tasked either to make selections from a set of options given their attribute values, to determine threshold values to inform future selections, or to create new options and attributes~\cite{brumar2025typology}.
When these tasks are preformed collaboratively, maintaining a shared awareness of the options and values that collaborators select can reduce redundant effort and ideally accelerate progress toward a state of consensus~\cite{mahyar2014clip}. 

Collaborative itinerary and event planning is one manifestation of a decision-making scenario, one in which collaborators evaluate and make selections from a set of options.  
\cref{fig:teaser} presents an interface for evaluating and selecting flights between origin and destination airports on a world map, with blue circles representing airports, and viable flights between them represented as purple arcs.
Such an interface could be used by a pair of family members or friends to plan a travel itinerary, or by the planning team of a conference or event, so as to identify venues that are logistically viable. 
As with all of the example interfaces to follow in this section, this interface allows a local collaborator to see their remote peer's intentions as well as their peer's reactions to their own interactions with the interface.

In \cref{fig:teaser}a, \userd{} \textit{points} at Madrid with their left index finger and at Stockholm with their right index finger, ephemerally selecting both locations and revealing a flight path between them.
This particular interface associates left-hand selections with origin locations and right-hand selections with destination locations, and these selections are reflected in respective list widgets composited in the top left corner of the interface. 
After observing this selection behavior, \usera{} is ready to evaluate some options with \userd{}, however this will require some negotiation and panning of the map via the \textit{bimanual grab} gesture (see video). 
In \cref{fig:teaser}b, \userd{} performs a \textit{spread} gesture with their left hand to make an ephemeral coarse selection of five candidate origins in Western Europe, while \usera{} performs the same gesture with their right hand, resulting in a similar coarse selection of three destinations in India.
Given this coordinated selection of origins and destinations, a list widget containing candidate flights now appears to the left of the composited interface, along with histograms reflecting the cost, travel time, and departure dates of these flights below them.
Finally, in \cref{fig:teaser}c, \userd{} uses their left hand to perform the \textit{pinch-and-move} gesture over this widget, which scrolls the list of flights. During this scroll, \userd{} spots an affordable flight leaving on an acceptable date between Madrid and Delhi, pointing toward it, but without activating its ephemeral selection. \usera{} maintains their coarse selection within India while using their other hand to ephemerally select the flight of interest.

\subsection{Exploratory Analysis}
\label{sec:apps:analysis}

Collaborative visualization scenarios involving two people is sometimes characterized as \textit{paired analytics}~\cite{arias2011pair}; like paired programming, this could include situations where a pair of visual analysts work together, or when a visual analytics application expert works alongside a subject matter expert, translating the latter's domain questions into interactions with an interface.

Two example interfaces speak to this scenario. 
First, in \cref{fig:apps:eda} (Scene 1), \usera{} and \userd{} take on the roles of political scientists analyzing the voting patterns of elected officials with respect to proposed legislative bills.
\usera{} asks whether the voting patterns of each party's establishment contingents are similarly homogeneous; in response, \userd{} performs a \textit{spread} gesture with their right hand to make an ephemeral coarse selection of moderate legislators in the right party, revealing their votes for and against the ten bills, represented as green and red links, respectively.

As a second example, in \cref{fig:apps:eda} (Scene 2) 
\usera{} and \userd{} are social scientists analyzing trends in inter-state migration patterns. 
As in the decision making scenario above, this interface shows data on a map and distinguishes a selection based on which hand performs the gesture. 
Here, \userd{} performs a \textit{spread} gesture with their right hand, making an ephemeral coarse selection of the Southeast states as migration destinations, painting them blue and revealing their largest sources of migration as arcs that transition from origin states (red) to destination states (blue).
Meanwhile \usera{} expresses an interest in recent migration patterns, \textit{pointing and tapping} to persist a selection in a radio widget composited in the top left, filtering the migration data to reflect the 2020s. 
Their coordinated selection reveals a ranked list of state-to-state migration to the Southeast during this period.

\subsection{Tutoring}
\label{sec:apps:tutoring}

One-on-one peer tutoring scenarios, particularly in science, technology, engineering, and mathematics (STEM) fields, often involve the use of visualization artifacts, diagrams, and other visual abstractions to communicate concepts. 
One inspiration for this scenario is a particular instantiation~\cite{perlin2017chalktalk} of Perlin~\etal's Chalktalk~\cite{perlin2018chalktalk}, a sketch-based presentation tool in which pen-based gestures are used to draw and control dynamic physical and mathematical models; in this demonstration~\cite{perlin2017chalktalk}, Perlin sketches using a light pen, compositing the sketches over his webcam video as he explains and animates them.
We anticipate that our approach could be similarly employed, albeit without requiring a specialized pointing device.
Moreover, our approach ensures a mutual visibility between tutor and student, allowing the former to ask questions about visual elements via deictic gestures. 
As for shared gestural control of composited visual aids, the tutor might selectively grant interactive privileges to the student to evaluate their learning.

Two of the implemented interfaces reflect this scenario.
First, in \cref{fig:apps:tutoring} (Scene 1), \usera{} is teaching \userd{} about graph topologies using a node-link representation.
He illustrates the shortest path between two nodes by performing a \textit{point} gesture with both index fingers, which ephemerally highlights the path between them and lists the nodes along the path in the left margin of the interface. 
To better see the connecting node along this path, \userd{} \textit{pinches and moves} the node to more clearly see the highlighted path. 
Second, in \cref{fig:apps:tutoring} (Scene 2), \usera{} is illustrating pairwise relationships between attributes in a tabular dataset using a scatterplot matrix (SPLOM). 
When they perform a \textit{spread} gesture over a set of data points in one of the scatterplots, they ephemerally highlight the same data points in the other scatterplots while dimming out unselected data points, drawing \userd{}'s attention to the different correlation relationships between attribute pairs. 

\subsection{Technical Interviewing}
\label{sec:apps:interviewing}

Lastly, our approach could also be used in technical assessments of job applicants, such as in system design and data analysis interviews.
\autoref{fig:apps:interview} (Scene 1) presents an example of the former, in which \usera{} provides \userd{} with a dynamic shared whiteboard interface, one where \userd{} can \textit{pinch and move} architectural components from a component menu on the left, generating clones of them and placing them where they deem appropriate. 
After placing them, \userd{} can perform \textit{pinch and hold} gestures on any two placed elements to add a visual connection between them. 
At any point, \usera{} can interject with \textit{point} gestures of their own, ephemerally highlighting elements, asking \userd{} to explain the rationale for their architectural design choices.
Meanwhile, \autoref{fig:apps:interview} (Scene 2) presents an example of the latter, in which \usera{} tests \userd{}'s visualization literacy by providing him with a Sankey diagram depicting the propagation of energy from natural sources to points of consumption. 
Here, \userd{} \textit{points} at two nodes in the diagram to ephemerally highlight the proportion of natural gas passing through the electric grid.

\section{Evaluation}
\label{sec:evaluation}

\begin{figure*}[h]
    \centering
    \includegraphics[width=1\textwidth]{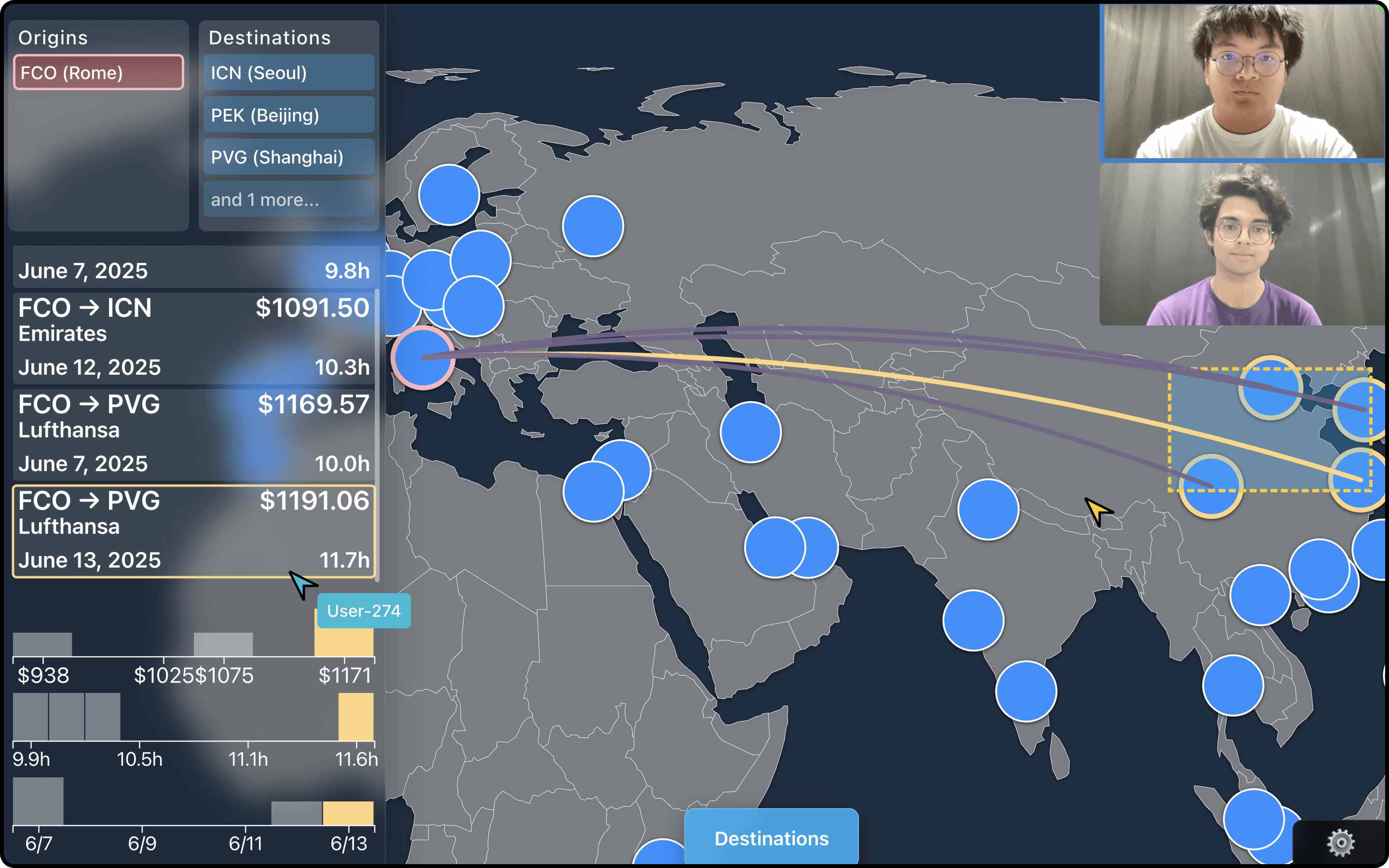}
    \caption{A screenshot of the baseline flight search application used in the evaluation. Remote collaborators can jointly control the interface via mouse interaction and see each other's cursor position.}
    \Description{A screenshot of the baseline flight search application used in the evaluation. Remote collaborators can jointly control the interface via mouse interaction and see each other's cursor position.}
    \label{fig:evaluation:baseline}
\end{figure*}

We evaluated our approach in a study designed around the \textit{decision-making} scenario described above in \cref{sec:apps:decision} and depicted in \cref{fig:teaser}. 
Our baseline for comparison was a collaborative visualization application (\cref{fig:evaluation:baseline}),
one in which the local and remote webcam feeds appear as thumbnail videos in the top right corner of the interface, replicating a conventional videoconferencing experience. 
However, this baseline application diverges from the convention of one-way screen sharing, a convention that we deem to be unfairly disadvantaged relative to our approach. 
Instead, our baseline application allows for shared mouse cursor awareness and control of visualization and interface widgets by either collaborator using mouse input, similar to previous work on synchronous collaborative visualization~\cite{neogy2020representing,schwab2020visconnect,han2024deixis}.

As a result, both baseline and Glass Chirolytics applications satisfy design goal \dii{} (\textit{support the collaborative analysis of complex data abstractions}; see \cref{sec:design:goals}), whereas only the latter supports \di{} (\textit{provide a common perspective on shared visual aids while coordinating face-to-face conversation}) and \diii{} (\textit{synchronize interaction with a gestural vocabulary commensurate with analytical use cases}).

The University of Waterloo research ethics board approved this study.

\subsection{Participants}
\label{sec:evaluation:participants}

We recruited 16 participants, a group reflecting diversity in terms of gender and age from the student and professional population in the locality of our institution. 
We were permitted to advertise the study via university mailing lists and on physical poster boards around campus, spanning multiple faculties. We did not specify explicit inclusion or exclusion criteria, such as having specialized knowledge about network data or prior data analysis experience.
As a result, the recruited participants reflect a range of education and prior experience with respect to gestural interfaces.
Eleven participants were between the ages of 21 and 25 and five were between the ages of 26 and 40. 
Eleven participants had an academic or professional background in computer science, three had a background in engineering, and two had a background in environmental science. 
14 participants used videoconferencing applications at least weekly. 
Nine participants reported having a limited experience with gestural interaction, while five reported a moderate level of experience, and two reported extensive experience. 
All of the participants reported having experience with visualization and analytics applications, with eight reporting a moderate level of experience and eight reporting extensive experience.
The study took an hour to complete, and we remunerated participants with a \$20 CDN multi-retailer gift card. 

\subsection{Setting, Apparatus, \& Procedure}
\label{sec:evaluation:procedure}

\begin{figure*}[h]
    \centering
    \includegraphics[width=\textwidth]{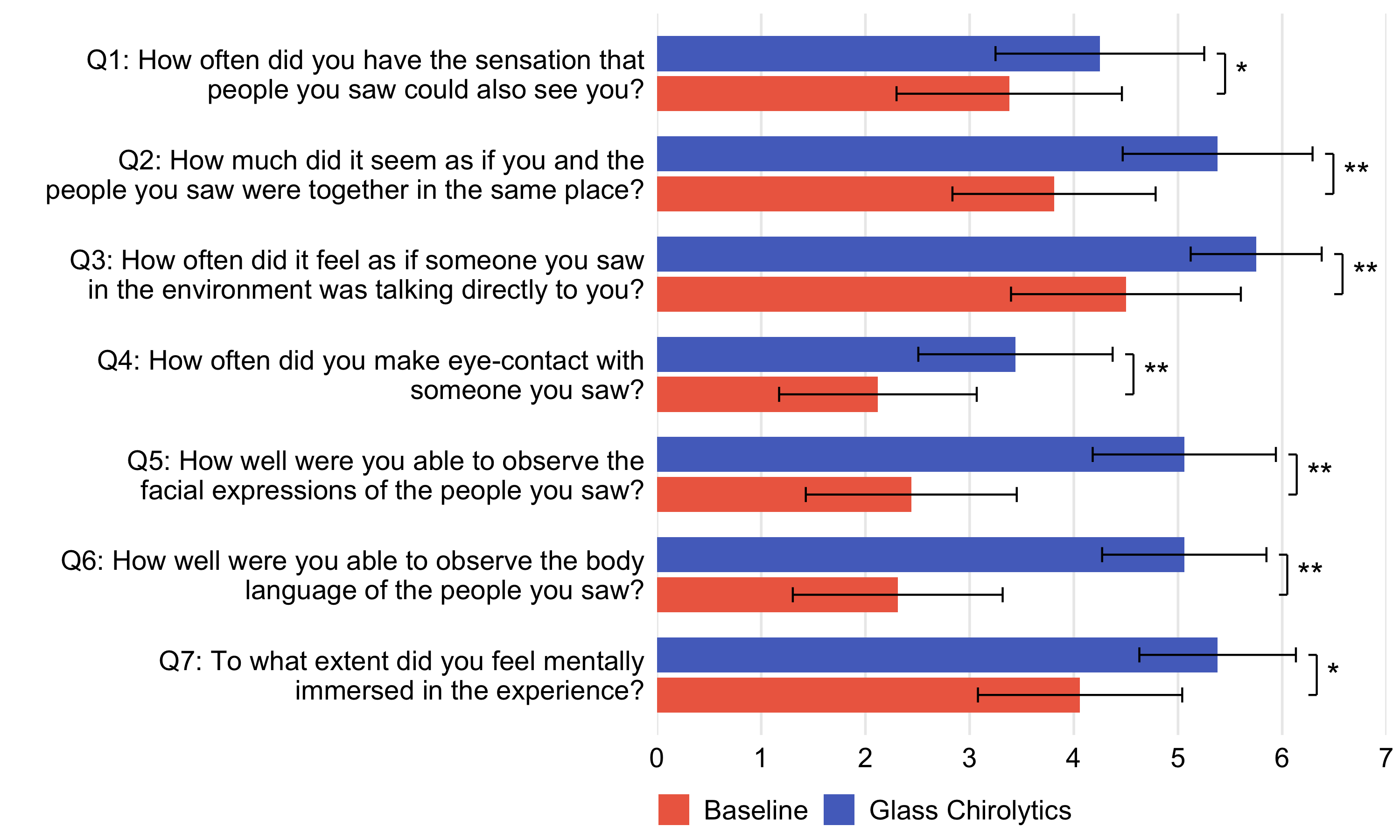}
    \caption{Temple Presence Inventory (TPI)~\cite{lombard2009tpi} presence scores for the baseline and Glass Chirolytics applications. $(* : p < .05, ** : p < .01, *** : p < .001)$. Error bars indicate 95\% confidence intervals.}
    \Description{A bar chart which contrasts the average Likert-scale responses for each of the seven Temple Presence Inventory questions between both applications in the study. For each question, two bars are shown depicting the average response value for both applications. All questions exhibit statistically significant different responses between the applications, with greater values shown for Glass Chirolytics.}
    \label{fig:tpi}
\end{figure*}

\begin{figure*}[h]
    \centering
    \includegraphics[width=\textwidth]{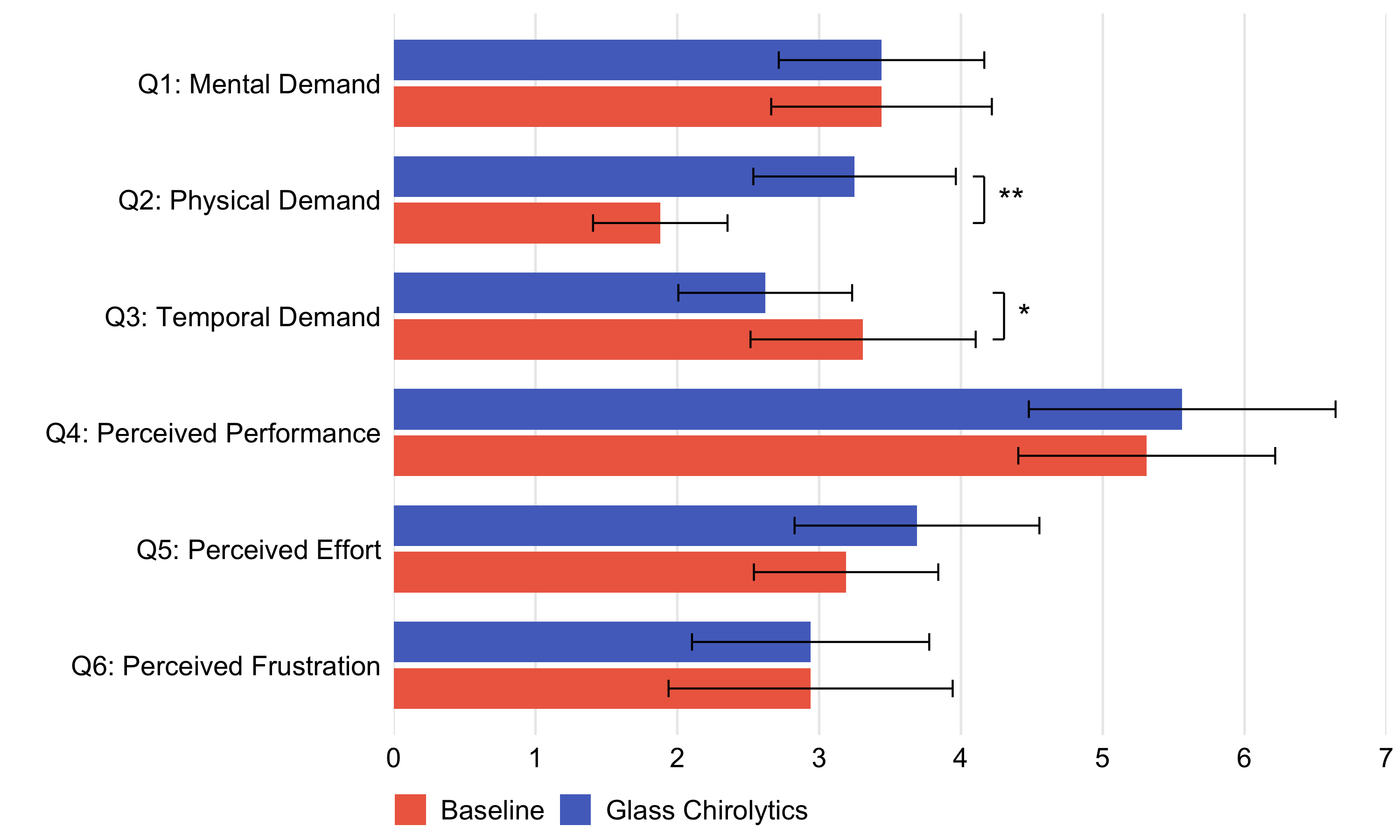}
    \caption{NASA-TLX~\cite{hart1988nasatlx} workload scores for the baseline and the Glass Chirolytics applications. $(* : p < .05, ** : p < .01, *** : p < .001)$. Error bars indicate 95\% confidence intervals.}
    \Description{A bar chart which contrasts the average NASA-TLX workload scores between both applications in the study. For each question, two bars are shown depicting the average response value for both applications. Physical demand and temporal demand have statistically significant differences. Physical demand is greater and temporal demand is lesser for Glass Chirolytics in comparison to the baseline.}
    \label{fig:nasatlx}
\end{figure*}

We conducted each study session in a controlled lab setting, with two participants per session. 
Upon arrival, participants signed a consent form, completed a questionnaire on demographics and prior experience with teleconferencing, gestural interaction, and visualization, and received a brief introduction to the study's purpose and procedures. 
To emulate a remote videoconferencing experience, we asked participants to sit at desks in separated partitions of the lab space. 
We provided each participant with an M1 MacBook Air laptop connected to a 27" monitor perched on an adjustable stand, a USB mouse, a 1080p Logitech USB webcam positioned atop the monitor, and a Logitech combined headphone-microphone USB headset.
The physical separation between participants ensured that they could not directly see or hear each other beyond the videoconferencing experience. 

The task that we assigned each pair of participants mirrors the \textit{decision-making} scenario described in \cref{sec:apps:decision}.
We selected this scenario as its visualization is representative of a high level of visual complexity relative to the other scenarios, reflecting highly interconnected data and featuring multiple coordinated views, with tasks necessitating a greater employ of our gestural vocabulary. 
Also unique among the others, this scenario required no specialized domain expertise, thereby avoiding situations in which there is an asymmetry of prior subject matter knowledge between the two participants.

The participants used an interactive flight search interface to arrive at a consensus travel decision. 
To achieve this, the participants role-played as a pair of travelers planning a trip to a shared destination that accommodated their individual constraints.
We indicated a small set of travel origins (\eg~New York City, Washington D.C.) as well as a deliberately vague \textit{cognitive region}~\cite{montello2003regions} as the travel destination, such as \textit{Western Europe} or \textit{Southeast Asia}, so as to promote deliberation between participants regarding the bounds of this region, from which they would narrow down to specific destinations (\eg~Paris, Bangkok).
We also gave each participant a unique set of travel constraints encompassing a budget, a departure date window, and a set of preferred airlines. Solutions for the task were flight itineraries that reflected the intersection of constraints given to both participants.
For this study, we procedurally generated a dataset containing two thousand flights, and ensured that both participants' constraints overlapped in such a way that would lead them to identify up to three solution itineraries satisfying their constraints. The task ended either after 15 minutes or when the participants had found all of the solution itineraries. 

Participants performed this task once with the baseline application and once with the Glass Chirolytics application.
We counterbalanced the ordering of the applications as well as the assignments of origins, destinations, and constraints across all participants. 
Prior to using each application, the researcher provided a tutorial consisting of a short prerecorded video and an interactive training session for practicing how to control the interface via mouse or gesture; we include the former as supplemental material. 

We recorded conversation audio as well as a video of the applications from both participants' perspectives during the study.
After using either the baseline or the Glass Chirolytics application, we asked participants to complete the Temple Presence Inventory (TPI)~\ \cite{lombard2009tpi}, an assessment used in prior remote collaboration research (\eg~\cite{qian2024chatdirector}) that reflects impressions of a collaborator's presence.
We also asked participants to complete the NASA-TLX~\cite{hart1988nasatlx}, which reflects subjective mental workload. 
For both instruments, participants answered with respect to a seven-point Likert-scale. 
Finally, the researcher conducted an open-ended group interview with both participants to gather qualitative feedback on their experience with both applications and their suggestions with respect to other application scenarios.

\subsection{Results: Performance, Presence, \& Workload}
\label{sec:evaluation:quant}

\bstart{Performance}
Each pair of participants identified at least one solution itinerary before the time elapsed in both conditions, and we observed no significant difference in terms of the number of solutions found. Due to the existence of multiple solutions and varying levels of rapport and conversation between participants (the experimenter encouraged participants to deliberate), the time taken to identify solutions is impractical to directly measure and compare.

For the TPI and NASA-TLX, we analyzed responses to the Likert-scale questions using the Wilcoxon Signed-Rank Test~\cite{wilcoxon1992individual}.
Following recommendations from previous research~\cite{siegel1957review}, we ensured that the sample size for this test exceeded 15 pairs. 

\bstart{Presence}
\cref{fig:tpi} shows the distribution of scores across the TPI~\cite{lombard2009tpi} questions.
Overall, we found that participants reported a significantly higher sense of presence with the Glass Chirolytics application relative to the baseline application across all seven questions ($p < 0.05$), encompassing mutual visibility, eye-contact, observation of facial expressions and body language, and a sense of immersion. 

\bstart{Workload}
\cref{fig:nasatlx} shows the distribution of scores across the NASA-TLX~\cite{hart1988nasatlx} questions. 
Relative to the baseline application, we found that participants reported a significantly lower level of temporal demand ($p = 0.0454$) and a significantly higher level of physical demand ($p = 0.002$) with the Glass Chirolytics application.
We revisit the significantly lower perceived temporal demand below in our analysis of video observations and interview responses.
We did not find different levels of reported mental demand, perceived performance, perceived effort, or perceived frustration. 

\subsection{Results: Video Observations}
\label{sec:evaluation:video}
In the baseline condition, interaction was predominantly serial, where one participant would monopolize interacting with the application while the other assumed a spectator role. 
For instance, P5 performed nearly all interactions while P6 watched and provided feedback; we observed a similar dynamic between P7 and P8 and between P11 and P12. 
Pairs often resorted to territorial strategies to avoid cursor conflicts, such as P3 explicitly asking P4, \textit{``do you want to hover over airports while I scroll on the list?''}, effectively assigning distinct interface areas. 
Furthermore, turn-taking relied on explicit verbal coordination, leading to awkward transitions and apparent friction. P13 and P14, for example, negotiated control using phrases like \textit{``let me do it,''} referring to list navigation, or \textit{``you go, you go''} when changing destinations.

In contrast, the Glass Chirolytics interface appeared to foster more fluid and simultaneous collaboration where rigid roles were less apparent. 
P5 and P6, who interacted serially with the baseline application, worked together simultaneously, with P6 moving the map in anticipation of a destination change while P5 scrolled the list. 
The embodied nature of the approach served as an inherent collision avoidance mechanism, reducing the need for territoriality and encouraging participants to work closely in the same screen space. This allowed P3 and P4 to take turns, quickly hovering over potential destinations in the same region, stopping only when they identified a destination that had a favorable flight distribution. 
Instead of a verbal coordination of turn-taking, we observed turn-taking initiated via gestural interjections and nonverbal cues.
For example, unlike their baseline behavior, P13 and P14 worked together in the same geographic area, using a combination of their partner's hand position, gaze, and conversation context to determine how to avoid collisions.
Altogether, the parallelization of interaction and reduced verbal coordination effort observed when participants used the Glass Chirolytics interface may account for the perception of reduced temporal demand mentioned above.

\subsection{Results: Interview Responses}
\label{sec:evaluation:qual}

We transcribed the open-ended group interview audio recordings using Microsoft Word's transcription tool, cleaned the resulting transcripts manually, and added our observations based on the study task video and audio recordings.
Given this dataset of utterances and observations, we conducted a reflexive thematic analysis \cite{braun2006thematic}.

\bstart{Glass Chirolytics provides a shared awareness of analytical intent and provenance}
A recurring perceived benefit of our approach was that it provided foresight into their partner's analytical intent via their gaze and hand positioning, with P12 stating that they ``\textit{could easily see where} [their peer] \textit{would be pointing and even before she selects the origin or the destination,}'' whereas in the baseline application, ``\textit{only after she selects one of them, I could see which one she selected.}'' 
This foresight promotes guidance, with P10 remarking that by ``\textit{being able to see people's hand motions, you can actually see what they're trying to do and guide them, versus what it's like on a mouse:} [\ldots] \textit{you can't see anything.}'' 
Despite a mutual visibility of each other's cursor in the baseline application, participants found that this offered minimal utility beyond a mere position. 
P4 commented that a cursor ``\textit{feels insufficient to capture the entirety of my intentions.}'' 
From the observer perspective, P6 noted that in the baseline application, ``\textit{even though you could see their mouse, it was a lot less evident what they were about to do.}'' 

Performing and witnessing gestural commands also increased a sense of accountability, with
P5 stating that ``\textit{it's easier to track with gestures what you} [selected] \textit{because you remember.}'' 
In contrast, they cited an instance when using the baseline application: ``\textit{we clicked on two flights unexpectedly\ldots and then we had to find out which one was wrong.}''
This apparent embodied sense of interaction provenance made it easier to identify and rectify mistakes.

Beyond gesture, the visibility of nonverbal cues like gaze and facial expression also improved communication flow according to several participants (P3, P8, P9). 
P9 cited an instance in which their peer was momentarily distracted while reading their assigned travel constraints offscreen, and in this moment P9 waited for their peer to stop reading before resuming the task.
Overall, participants generally paid less attention to each other when using the baseline application, with multiple individuals (P3, P11, P12) reporting to have never looked at the other participants' thumbnail video feed.

\bstart{Shared gestural control: potentially distracting at first, but quickly learned and even deemed to be `fun'}
While most participants quickly gained proficiency with the gestural vocabulary, some remarked that gestures detracted from the conversation, at least initially. For instance, P4 reflected to P3 that at when first using the Glass Chirolytics interface, ``\textit{I had not enough brain power to observe you.} [\ldots] \textit{I was trying to get used to the hand motions.}'' 
Other participants noted a need to be continuously cognizant of the camera as well as the gesture classifier and its limitations throughout the task.
P9 disliked having to ``\textit{make sure my hand is always flat and facing the camera}'' when interacting with the application, wishing it was ``\textit{able to identify even if your hands} [are] \textit{off axis.}''

Despite any initial difficulties with the gestures, most participants were able to quickly learn the gestures and were proficient enough to complete the task, despite the brevity of the training exercises before the task commenced.
Discussing the gestural learning curve, P8 commented that it was ``\textit{not very big}'' and took around ``\textit{two to three minutes\ldots after that, it becomes easier.}''
P9 told us that learning how to use the gestural interface \textit{``was easier than it seemed,''} whereas with the baseline interface, they said: \textit{``I thought would be easy, but I actually found it more difficult.}''
Participants reported learning by observing their peers, as reflected on by P10: ``\textit{I got to see him do it first, so it made it a little easier seeing someone else do it.}''
Similarly, several participants (P1, P13, P14, P16) described the experience of using this interface as being more fun, playful, and enjoyable than the baseline interface.

\bstart{Reactions to shared gestural control suggest a rich interaction design space}
Multiple participants (P5, P6, P8, P14) stated that they found gestures to be more intuitive than using a keyboard and mouse. 
For instance, P5 and P6 agreed that using their left and right hands for selecting origins and destinations, respectively, felt more natural than mode-switching with a mouse and keyboard. 
P8 would similarly characterize mouse- and keyboard-based mode-switching as abstract, while having the modes embodied in the hands felt like a tangible control scheme. 

Discussion in the open-ended group interviews often turned to expanding the gestural vocabulary.
One common suggestion (P1, P9, P11, P12, P13) was the ability to draw annotations with the hands. 
P1 suggested repurposing the \textit{spread} gesture; instead of coarsely selecting, this gesture could spawn a circular annotation.

Participants also bemoaned the lack of an idle state in which no gestures are recognized: a mode in which one can keep their hands within the camera's view without inadvertently triggering changes to the visualization interface, such as by scratching one's head (P6). 
To this point, P5 requested ``\textit{a way to filter out other hand movements and then catch only when you mean to use the system,}'' suggesting a potentially automated means of detecting interactive intent.

Other participants suggested expanding the gestural vocabulary by using gestures in tandem with other modalities.
For instance, P2 and P15 suggested overloading gestures with multiple commands associated with different modes (\eg~idle, selection, navigation, annotation), and that this mode-switching would be done with the mouse or keyboard.
Meanwhile, P16 suggested voice as an additional input modality, similarly complementing or modulating gestural input. 

Meanwhile, other participants suggested that the gestural vocabulary need not be limited to one's hands.  
Citing large wall-mounted displays, P3 and P4 proposed incorporating the entirety of one's body, with P4 remarking that ``\textit{half of my body is kind of useless and half of my body is not,}'' reflecting an interest to control the interface with their whole body, such as by walking toward and away from different parts of the display.  

\bstart{Participants suggested other collaborative scenarios involving complex data artifacts}
While our approach elicited a range of application scenarios, we focus on three scenarios involving visually complex data artifacts:

\istart{Codebases}
First, P2 described the challenges of remote code collaboration in Zoom meetings where one person shares their screen, and ``\textit{the other person doesn't understand what part} [of the code] \textit{you are you talking about.}'' 
They suggested that the ability to ``\textit{point out exactly to different places}'' may accelerate this work, particularly given the many alternative visual representations of code bases and system logs.

\istart{Tables}
Another suggestion from P2 reflects a need for data analysts to look at spreadsheets or tables together, an omnipresent category of artifact in data science workflows~\cite{bartram2021untidy}: ``\textit{we could even work in Excel} [\ldots] \textit{being able to see the other person and see how} [they] \textit{select different parts of the screen} [\ldots] \textit{I think that could be useful.}'' 

\istart{3D models}
P8 suggested applying the Glass Chirolytics approach to remote conversations involving a 3D model, where either participant could pull apart or reassemble parts of it via coordinated gestures.
They also envisioned applications in educational settings, 
such as those involving complex engineering drawings or anatomical models,
where an intent is either to teach or learn about part-whole relationships.
\section{Discussion}
\label{sec:discussion}

By reflecting on the Glass Chirolytics approach and our study results, we now present design implications for tools that aim to support synchronous and remote collaboration around data, propose new research directions, and discuss the limitations of our implementation and evaluation.

\subsection{From Design Decisions to Design Implications}
\label{sec:discussion:implications}

First, we revisit and extend our key decision decisions (\di{} --- \diii{}: \cref{sec:design:goals}) into design implications for future collaborative visualization applications. 

\bstart{Extend the mutual awareness of analytical intent with additional visual cues}
Participants in our evaluation valued the visibility of nonverbal cues prioritized by our approach.
Like in Tang~\etal's projection of remote collaborators' hands over a tabletop display ~\cite{tang2010threes}, our approach also appeared to give collaborators insight into their remote peers' intents, insight that may lead to fewer collisions, such as simultaneous attempts to manipulate the same element.
Given this awareness, it may not be necessary to employ locking mechanisms on interface elements, such as in cursor-based collaborative visualization approaches~\cite{schwab2020visconnect}.
Nor would it be necessary to highlight recently manipulated elements, such as with ephemeral afterglow effects~\cite{baudisch2006phosphor}, because the remote collaborator's hand movements already accomplish this. 

While we may not need to draw further attention to a remote collaborator's hands, it nevertheless may be fruitful to bring awareness to other nonverbal cues.
Given our study participants' hesitation at times to use their hands for fear of unintentional interaction, we could track and communicate a remote collaborators' gaze~\cite{koch2025multimodal}, so as to determine if they are looking at the same elements as the local collaborator.
Going further, we could recognize a remote collaborator's facial expressions as a means of inferring their affective state~\cite{murali2021affectivespotlight}, which might be particularly helpful to communicate to a tutor, interviewer, or anyone aiming to reach a consensus decision with their peer.
However, returning to \di{}, any accentuated nonverbal cues or inferences made from them should not detract from face-to-face conversation. 

\bstart{Accommodate visualization artifacts of further complexity}
Across our scenarios (\cref{sec:apps}), the visualization elements reflected multidimensional, relational, and (geo-)spatial data abstractions. 
Although this scope might satisfy many collaborative visual analysis scenarios, truly satisfying \dii{} means accommodating greater complexity. 
For two-dimensional visual representations, either collaborator should be able to load or paste any hierarchical SVG-based object on to the shared `glass'. 
However, a complication creeps in when we begin to consider rich spatial data abstractions represented visually as 3D volumes, such as simulation models (as suggested by study participant P8) or isometric-perspective topographical maps. 
With 2D visuals, the reciprocal compositing over mirrored video produces a shared perspective, whereas with 3D visuals, a remote collaborator will appear as though they have a different perspective, thereby conflicting with \di{}.
Despite this complication, we expect that aspects of Glass Chirolytics could be applied in collaboration around 3D visualization artifacts. 
At a minimum, we could accept that the local and remote participant will have different perspectives, looking at 3D content from opposing sides, such as in the Spatialstrates collaborative mixed reality ecosystem~\cite{borowski2025spatialstrates}.
Alternatively, we could allow for toggling between multiple perspectives, akin to the Blended Whiteboard approach~\cite{gronbaek2024whiteboard} for mixed reality collaboration, in which a face-to-face orientation facilitates rapid turn-taking in brainstorming, presenting, and sensemaking, while a side-by-side orientation ensures a common perspective on the content.
The latter orientation would likely require compositing, segmenting, and juxtaposing video input from both participants.
Future work could therefore study how participants freely toggle between orientations when performing open-ended collaborative analysis tasks with complex representations of spatial data.

\bstart{Support the journey from exploration to communication and back again}
Informal collaborative visualization is bound to alternate between exploration and explanation, between data analysis and storytelling with data~\cite{gratzl2016visual}.
This alternation might be particularly evident in pair analytics~\cite{arias2011pair} scenarios in which there is role asymmetry between analyst and domain expert. 
A question this elicits is whether our gestural vocabulary, motivated by analytical use cases (\diii{}), should be expanded to recognize and act on the rhetorical and expressive gestures associated with persuasive data presentations~\cite{hall2022chironomia}.

A larger gestural vocabulary might compromise the learnability of our approach, although it is possible, based on our study participants' comments, that watching others perform these gestures may accelerate learning.
In particular, a local collaborator watching a remote peer who is more experienced with the gestural vocabulary may learn at a faster pace than our study participants, who were on equal footing at the beginning of the study. 
Nevertheless, if the gestural vocabulary of our approach is to be expanded, the temptation to introduce mode-switching (\eg between idle, analysis, and storytelling modes) via keyboard or mouse commands is problematic because these mode switches are not likely to be visually apparent to remote collaborators in the same way that hand gestures are.
Alternatively, recent work suggests that in lieu of a fixed gestural vocabulary, the meaning of gestures could be inferred in real time with approaches like the LLM-based GestureGPT~\cite{xin2024gesturegpt}, which links untrained gestures to predicted interface actions.

Our gestural vocabulary thus far maps to selecting elements, reconfiguring their placements, and exploring their spatial distribution, with some selection gestures triggering filter events and the generation of connections between elements. 
Other common interactions with conventional desktop visualization applications~\cite{yi2007toward} do not have obvious gestural counterparts, such as changing the mapping of data attributes to visual variables, or changing the level of abstraction or detail. 
For these interactions, we might look to additional modalities such as speech input~\cite{srinivasan2023voice}, sketch input~\cite{perlin2018chalktalk}, or interaction with tangible icons~\cite{takahira2025tangiblenet} to complement our existing set of deictic gestures.

We remarked in \cref{sec:apps} how the scope of our approach focuses on synchronous and collaborative analysis and decision-making activities, and that these episodes form part of a larger fabric of data work, much of which being conducted asynchronously and individually.
This invites the question of whether and how the scope of our approach could be expanded to support activities that precede our scenarios, such as data preparation and shaping interfaces or visualization and dashboard design, activities in which having another set of eyes (and an another pair of hands) could have downstream benefits.

\subsection{Limitations}
\label{sec:discussion:limitations}

An obvious limitation of our approach is that it is intended for face-to-face conversations between two people with the same role, and thus it is unclear whether it is also suitable for small groups, such as a team of data scientists~\cite{brehmer2022jam}.  
Solutions could involve relegating all but one or two members of the group to a passive status, in which they see a side-by-side compositing of the active speakers. 
Alternatively, we might adopt an approach like MirrorBlender~\cite{gronbaek2021mirror}, in which participants can scale and position their video feeds in a unified composited display; however, this approach may limit the reach of any one group member's gestural control, as well as the mutual visibility of these gestures and other nonverbal communication. 
Expanding to larger groups may necessitate other forms of awareness of remote collaborator activity, such as via aggregated visual annotation layers on composited visualization and interface elements~\cite{chung2021livestream}.
A limitation with this interaction approach is the lack of precision in mid-air hand gestures relative to mouse and keyboard input; as a consequence, interactive elements must be sufficiently large to ensure reliable manipulation.

Our evaluation also has four noteworthy limitations. 
First, while one or more of the four scenarios described in \cref{sec:apps} are likely to be familiar to those with STEM education and work experiences, we opted to focus on a single scenario of collaborative decision making in our study, a scenario that may not occur with similar frequency for all of our participants. 
Second, we did not explicitly seek out participant pairs with existing levels of rapport and trust, and we acknowledge that these factors may impact the interpersonal awareness of nonverbal cues and collaborative working styles. 
Third, given the length of our evaluation and the time spent working with the Glass Chirolytics interface, we are unsure how fatiguing it would be to engage in longer conversations supported by our approach.
Finally, the individual gestures in our vocabulary (\cref{sec:design:gestures}) were not systematically evaluated with respect to their accuracy, precision, or memorability relative to alternative candidate gestures, and thus further experimental study is warranted.

\subsection{Future Research Opportunities}
\label{sec:discussion:fw}

\bstart{Report on longitudinal use by analysts and tutors}
We are eager to better understand any long-term benefits of our approach in scenarios spanning multiple usage sessions, such as exploratory data analysis and STEM tutoring. 
In particular, we call for investigations into whether this approach improves comprehension and information retention, and whether the dynamics and metrics of collaboration change over time. 
The impact of role and expertise asymmetries would also be worthy to study longitudinally. 
Meanwhile, in collaboration with cognitive scientists, we are curious to determine whether the observation of a conversation partner's noncommunicative and manipulative gestures has a learning benefit akin to what is exhibited with co-located communication through activation of the human mirror neuron system (hMNS)~\cite{dickerson2017role}.

\bstart{Investigate asymmetries of technology}
Another question is how the collaborative experience changes in cases where one collaborator casts their video and interacts via gesture while the other remains off-camera and interacts via standard input devices, or if they collaborate via avatar while wearing a head-mounted display~\cite{tong2023asymmetric}.
Collaborators with large displays~\cite{zillner20143dboard} and multiple webcams also introduce asymmetries.
For instance, those working from large rooms outfitted with large displays could interact via full-body gestures~\cite{saquib2019interactive}, as suggested by study participants P3 and P4. 
If their peers are limited to small displays and conventional distances from their cameras~\cite{bailenson2021nonverbal}, future adaptations of our approach must reconcile disparities in the scale of composited elements and the reach of gestures.
Unsurprisingly, our study participants reported that the physical demand imposed by the Glass Chirolytics interface was significantly higher than that of the baseline interface.
Using multiple webcams might alleviate this demand by positioning one webcam in a conventional position to capture the face and another positioned above the work surface in front of the monitor, pointing down to capture the hands gesturing on that surface.
This camera perspective has made for compelling data journalism with hand-drawn charts~\cite{fong2022}, which suggests that it could also be a viable perspective for compositing visual aids. 
Furthermore, this approach may represent a middle ground between conventional screen sharing and Glass Chirolytics, one that retains some benefits of nonverbal communication without inducing considerable levels of fatigue.

\bstart{Augment our approach with speech-generated visual aids}
Finally, our approach could be complemented by the addition of speech input.
Speech input could, for instance, be used to trigger mode-switching, particularly as mode-switching gestures expand the gestural vocabulary and may be distracting to the remote observer.
Beyond the aforementioned prospect of using speech to manipulate existing composited visual aids~\cite{srinivasan2023voice}, conversation speech could also be used to generate, retrieve, and recommend other context-appropriate visual aids.
With systems like Liu~\etal's Visual Captions~\cite{liu2023visualcaptions} or Xia~\etal's CrossTalk~\cite{xia2023crosstalk} bringing contextually-appropriate images and documents into the conversation, we are hopeful that similar approaches could be taken to generate contextually-appropriate, valid, and interactive data visualization artifacts for the shared glass.
\section{Conclusion}
\label{sec:conclusion}

We introduced Glass Chirolytics, an augmented videoconferencing approach designed to foster engaging and multimodal face-to-face analytical conversations involving complex data abstractions. 
It employs a reciprocal compositing of shared visualization artifacts over the mirrored webcam video of one's conversation partner, which we pair with bimanual mid-air gestures for manipulating visualization and interface elements. 
These design decisions bring collaborators' hands back into the conversation by virtue of them appearing in the video frame: a step toward restoring the benefits of nonverbal communication lost in  videoconference meetings relative to co-located meetings. 

We evaluated our approach with eight pairs of participants in a comparative study, in which participants completed travel itinerary decision-making tasks given a large flight schedule dataset, both with a Glass Chirolytics application and with a baseline application reflecting conventional videoconferencing with mouse control of a shared interface.
Our findings suggest that our approach significantly enhances a sense of presence while reducing the temporal demand of collaborative analysis. 
Participants' remarks also suggest that our approach provided an awareness of the analytical intent of one's conversation partner, an awareness that was lacking in the baseline experience.
Given these results, we are optimistic about the potential of gesture-controlled applications for enriching remote collaboration around data, and for accelerating learning in cases where there is an asymmetry of knowledge and tool experience between conversation partners.

\begin{acks}
We thank Anchit Mishra, Mohammad Abolnejadian, Jenny Zhang, Jessica Chen, and Skylar Ji for their feedback on the project. This research was supported by a University of Waterloo Cheriton School of Computer Science Undergraduate Research Fellowship and a Natural Sciences and Engineering Research Council of Canada
(NSERC) Undergraduate Student Research Award.
\end{acks}

\bibliographystyle{ACM-Reference-Format}
\bibliography{main}
\end{document}